\documentclass[aps,prb,twocolumn,amsmath,amssymb,superscriptaddress]{revtex4-1}

\usepackage{braket, units}
\usepackage{hyperref}
\usepackage{graphicx}
\newcommand{\Hopen}{H_{\mathrm{open}}}
\newcommand{\adj}{\mathrm{adj}\, }

\newcommand{\imag}{{\mathrm{imag}}}
\newcommand{\lbit}{{\text{l-bit}}}
\newcommand{\inter}{{\text{int}}}
\newcommand{\ff}{\text{finite-frac} }

\begin{document}

\title{Extracting many body localization lengths with an imaginary vector potential}
\author{Sascha Heu\ss en}
\affiliation{Institute for Quantum Information and Matter and Department of Physics, California Institute of Technology, Pasadena, CA 91125, USA}
\author{Christopher David White}
\affiliation{Condensed Matter Theory Center, University of Maryland, College Park, Md, 20742, USA}
\author{Gil Refael}
\affiliation{Institute for Quantum Information and Matter and Department of Physics, California Institute of Technology, Pasadena, CA 91125, USA}
\begin{abstract}
  One challenge of studying the many-body localization transition is defining the length scale that diverges upon the transition to the ergodic phase.
  In this manuscript we explore the localization properties of a ring with onsite disorder subject to an imaginary magnetic flux.
  We connect the imaginary flux which delocalizes single-particle orbitals of an Anderson-localized ring with the localization length of an open chain.
  We thus identify the delocalizing imaginary flux per site with an inverse localization length characterizing the transport properties of the open chain.
  We put this intuition to use by exploring the phase diagram of a disordered interacting chain,
  and we find that the inverse imaginary flux per bond provides an accessible description of the transition and its diverging localization length.
\end{abstract}

\maketitle


\section{Introduction}

A large body of evidence shows that one-dimensional fermionic quantum systems with both (local) interactions and sufficiently strong disorder will exhibit a cluster of traits known as \textit{many-body localization}\cite{basko_metalinsulator_2006,oganesyan_localization_2007,pal_many-body_2010}.
These include long-time memory of an initial state, conductance exponentially small in length, slow entanglement growth, and near-Poisson level statistics.

The localization length of single particle orbitals is a standard measure of localization.
This notion applies to interacting localized systems as well as noninteracting systems.
In fact, the localized nature of the many-body localized phase implies that it can be described by a so-called  \textit{$\ell$-bit} Hamiltonian\cite{serbyn_local_2013,huse_phenomenology_2014, huse_phenomenology_2014}.
The Hamiltonian can be written in terms of mutually commuting single-particle occupation operators $\tilde n_j$ with local support as
\begin{align*}
  H = \sum E_j \tilde n_j
  &+ \sum_{jk} J^{(2)}\tilde n_j\tilde n_k\\
  &+ \sum_{j_1 j_2 j_3} J^{(3)}_{j_1j_2j_3}\tilde n_{j_1}\tilde n_{j_2} \tilde n_{j_3}\;.
  &+ \cdots
\end{align*}
where the interactions $J^{(n)}$ are presumptively short-ranged.
Explicitly constructing the $\ell$-bits would fully solve the quantum dynamics of a many-body localized chain; the problem of doing so has attracted much attention.\cite{ros_integrals_2015,chandran_constructing_2015,khemani_obtaining_2016,obrien_explicit_2016,pekker_fixed_2017}

The first phenomenological and numerical treatments of the MBL transition tended to concentrate on entanglement and transport times,\cite{vosk_theory_2015,potter_universal_2015,zhang_many-body_2016,goremykina_analytically_2019}
and computed  the gap ratio, the entanglement entropy of eigenstates, or decay of local observables.
\cite{pal_many-body_2010,oganesyan_localization_2007,khemani_critical_2017,PhysRevB.91.081103,kjall_many-body_2014}
But the microscopic avalanche picture \cite{thiery_microscopically_2017,de_roeck_stability_2017,luitz_how_2017,thiery_many-body_2018} and recent renormalization analysis \cite{morningstar_renormalization-group_2019} hinge on the correlation lengths of perturbatively-constructed $\ell$-bits. 

The most direct way to probe many-body localization, however, should be through finding the appropriate localization length of single particle creation operators. Such a localization length, which is analogous to the Anderson localization scale, would be most relevant to transport-related questions. A localization length could  be defined from the support of the $\ell$-bits. But constructing $\ell$-bits is difficult and relies on variants of exact diagonalization, Wegner flow or matrix product state methods;
moreover, the same physical Hamiltonian can be described in terms of many different sets of $\ell$-bits.
Extracting localization length from $\ell$-bits of limited-size systems is therefore, difficult and ambiguous. Nonetheless, Refs. \onlinecite{Vadim-scales, Simon, Cappellaro}  succeed in extracting localization length by constructing $\ell$-bit operators and exploring their decay.

In this manuscript we show that exploring the response of a system to non-Hermitian hoppings (namely, to an imaginary flux) provides a direct way to address the definition of a localization length. Furthermore, it does not require knowledge of any $\ell$-bits properties, and addresses directly the localization length most relevant for transport. We introduce an imaginary vector potential, which maps to a ``tilt''---an asymmetry in the tunneling rates between neighboring lattice sites.
While at small tilts all eigenvalues of the system's Hamiltonian remain real, they develop imaginary parts at some critical tilt. 
We argue that the critical tilt of a non-Hermitian system probes the $\ell$-bit localization length $\xi_{\lbit}$ of its zero-tilt Hermitian limit, while the distribution of points at which successive eigenvalues develop imaginary parts (``exceptional points'') probes the $\ell$-bit interaction scale $J_\inter$. We show explicitly that the critical tilt in a ring (a chain with periodic boundary conditions) is the inverse localization length of the open chain with the same disorder realization. We use this to describe the many-body localization transition and extract its phase diagram. 

We first introduce our model in Sec.~\ref{s:model}.
We then articulate the connection between critical tilt in a ring and the localization length of an open chain in the single-particle case (in Sec.~\ref{s:sp});
in doing so,
we extend the work of Ref.~\onlinecite{shnerb_winding_1998} to connect the critical tilt on a ring to a Green's function on the \emph{open} chain.
We then consider generalizations to the many-particle, non-interacting case (Sec.~\ref{s:mp:nonint}) and to the interacting case (Sec.~\ref{s:mp:int}),
where we show a connection between the distribution of exceptional points and the l-bit interaction strength.
Finally, in Sec.~\ref{s:phase} we use the critical tilt to map the MBL transition of the isotropic random-field Heisenberg model.
We find a phase diagram and critical exponents broadly consistent with \cite{PhysRevB.91.081103}. In addition, we see that the  MBL regime is reentrant as a function of interaction strength. 



\section{Introducing Non-
Hermitian Hopping to the Many-Body Localization Model}
\label{s:model}

\subsection{Background}

The non-Hermitian hopping problem in a tilted disordered lattice was proposed as an effective model for vortex pinning in non-parallel columnar defects \cite{hatano_localization_1996,hatano_vortex_1997,hofstetter_non-hermitian_2004,affleck_non-hermitian_2004}. Indeed, the anhermiticity of the hopping operator represented the tilt of the columnar pinning defects relative to the external field. The critical tilt in this single particle problem was shown to be intimately related to the localization properties of the zero-tilt system.\cite{shnerb_winding_1998}
The relationship between critical tilt and correlation length has received much interest, as have various properties of the single-particle spectrum at fixed tilt. \cite{brouwer_theory_1997,brezin_non-hermitean_1998,feinberg_non-hermitean_1999,feinberg_spectral_1999,kolesnikov_localization-delocalization_2000,heinrichs_eigenvalues_2001,heinrichs_theory_2002}. Very recently, non-Hermitian tilt was also introduced to interacting systems \cite{hamazaki_non-hermitian_2019,panda_entanglement_2020}.

 \subsection{Model}
 
We study spinless fermions hopping on a one-dimensional lattice with a random onsite chemical potential and an imaginary vector potential.
The system's Hamiltonian is 
\begin{align}
  \begin{split}
    \label{eq:ham}
    H &= t \sum_j [e^gc^\dagger_j c_{j + 1} + e^{-g}c^\dagger_{j+1}c_j] \\
    &\qquad + U \sum_j n_j n_{j+1} + \sum_j h_j n_j\;.
  \end{split}
\end{align}
with the random onsite potential $h_j$uniformly distributed on $[-W,W]$.
We set the bare hopping to $t  = 1$.
When $U=2$, this model is Jordan-Wigner equivalent to the random-field Heisenberg model.
In Sec.~\ref{s:sp} we work in the single-particle sector;
in the subsequent sections we work at half-filling.

The Hamiltonian \eqref{eq:ham} is non-Hermitian, 
with anhermiticity parametrized by the imaginary vector potential, or ``tilt'', $g$. In an open chain the tilt $g$  could be removed by a similarity transformation 
\begin{equation}
S=e^{\sum\limits_j g j n_j}.
\label{similarity}
\end{equation}
In a ring, the imaginary flux cannot be removed by such a similarity transformation, and imaginary parts can appear in the energy eigenvalues of the system. 

The imaginary eigenvalues are directly related to delocalization on the ring.  
For $g > 0$, the system prefers leftwards hopping, but if $g$ is small one expects the system to remain localized.
Localized orbitals cannot explore the entire ring, and therefore their energy eigenvalues remain real.
At large $g$ the preferential leftward hopping dominates,
and one expects the system to be delocalized.
Orbitals that wrap around the ring necessarily develop an imaginary part to their energy.
In the next section we explain how to precisely relate the tilt $g$ at which imaginary parts appear to the localization length of orbitals.

\section{Critical tilt and localization length: the single-particle case}\label{s:sp}

\begin{figure}[t]
	\raggedleft
	\includegraphics[width=1.1\linewidth]{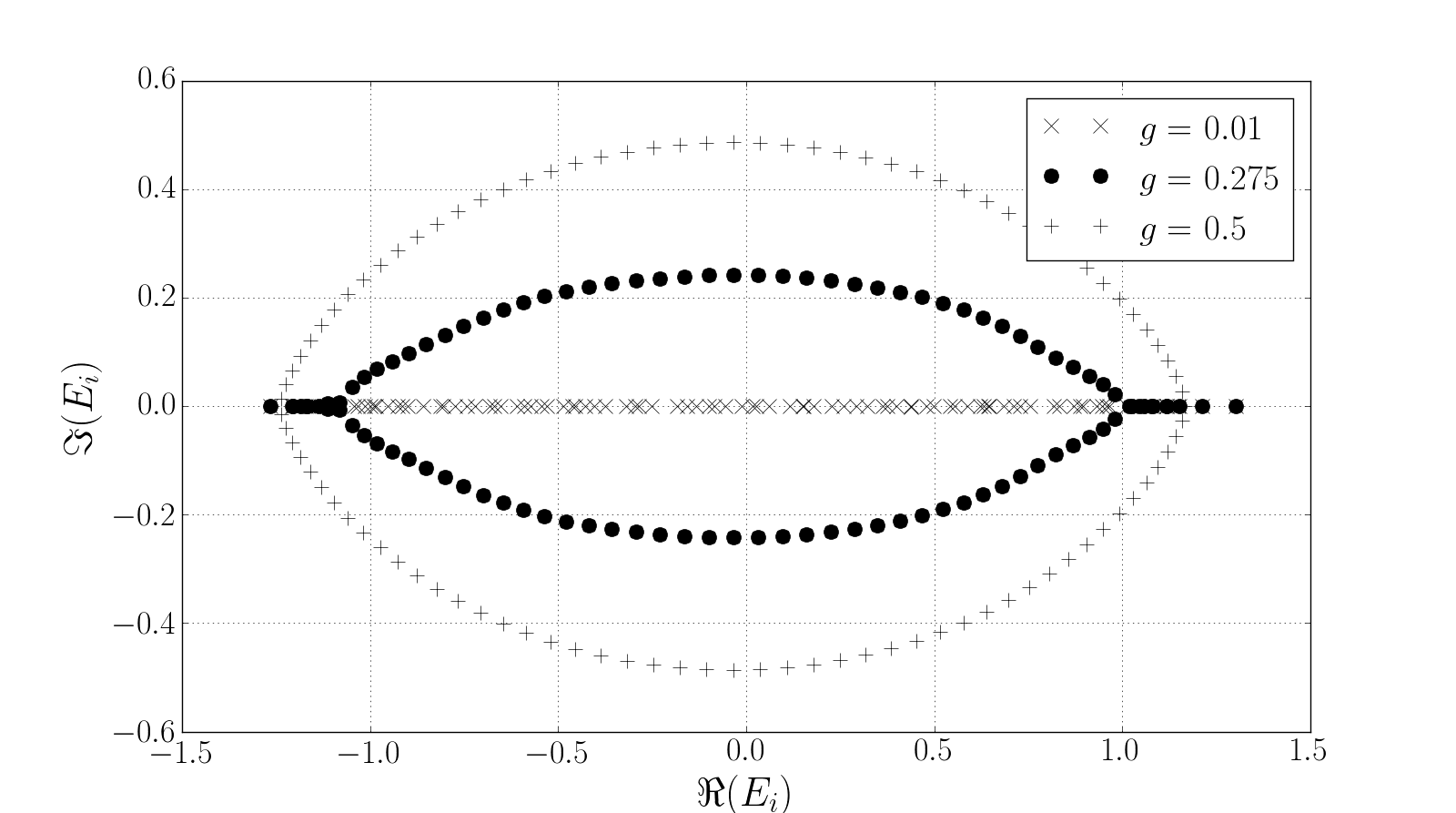}
	\caption{\textbf{Eigenenergies for one particle} on a chain of $N = 100$ sites with disorder width $W = 1$ at three tilts $g$.
          For sufficiently strong gauge field the single-particle spectra form ellipses emerging from the band center.
          }
	\label{fig:spec1} 
\end{figure} 

%
%

As the tilt in a ring increases, the Hamiltonian's eigenvalues develop imaginary parts.
The point at which an eigenvalue develops an imaginary part is called an \textit{exceptional point}.
We seek a precise relationship between the localization length of the open chain and the location of the exceptional point.

We can get some intuition for the processes involved and see what ingredients are required by a heuristic argument, in which we gauge the imaginary flux to one link and add that link perturbatively.
Consider the Hamiltonian of Eq.~\eqref{eq:ham} on $N$ sites with periodic boundary conditions, and take the single-particle case.
At $g = 0$ its single-particle eigenstates are exponentially localized, and, therefore, cannot explore the flux penetrating the ring.
An eigenstate $\ket n$ centered on some site $j$, for instance, can be asymptotically written as
\begin{equation}
  \ket n \sim \sum_{j'} e^{-|j-j'|/\xi} \ket {j'}.
\end{equation}
Through a similarity transformation as in Eq.~\eqref{similarity}, we can shift all of the imaginary vector potential to the far side of the system, away from $\ket n$'s center site $j$. The imaginary flux would then be shifted to $\bar j = j + N/2 \mod N$, and the Hamiltonian would be
\begin{align}
  \begin{split}
    H &= \sum_{j' \ne \bar j} [c^\dagger_{j'} c_{j'+1} + h.c.] + \sum_{j'}h^z_{j'} n_{j'}\\
    &\quad + e^{Ng}c_{\bar j}^\dagger c_{\bar j + 1} + e^{-Ng}c_{\bar j + 1}^\dagger c_{\bar j}\;.
  \end{split}
\end{align}
Imagine now adding the anhermitian hopping on the link $\bar j, \bar j + 1$ perturbatively.
  The perturbation becomes important when the resulting change in energy is comparable to some energy difference $\Delta E$ in the closed chain: that is, when
\begin{equation}
  \label{eq:intuitive-argument}
  1 \sim \bra n te^{Ng} c^\dagger_{\bar j + 1}c_{\bar j} \ket n \sim e^{N(g - \xi^{-1})}
  { \Delta E^{-1}}\;;,
\end{equation}
where $\Delta E$ is some energy difference in the closed chain.
Immediately we see that the anhermiticity is important when
$g \sim \xi^{-1}$.
This crucial insight---that the tilt competes directly with the localization properties of individual eigenstates---goes back to the work of Hatano and Nelson (e.g. Ref.~\onlinecite{hatano_vortex_1997}).
 But we also see the three ingredients that will be important in our detailed calculation: the tilt $g$, the end-to-end hopping matrix element in eigenstates of the open chain, and energy differences in the open chain.
  Although our detailed calculation in Secs.~\ref{s:sp:det}-\ref{s:sp:wol} applies only to single-particle (non-interacting) systems,
  we hope that this perturbative approach will in the future yield more precise insight into interacting (many-body localized) systems;
  we will return to it in interpreting our results for those systems.

For $g > g_c$, the eigenstates with complex energy eigenvalues resemble a plane wave. \cite{hatano_vortex_1997}
(Recall that we work in the single-particle sector.)
Therefore these eigenstates have (complex) energy $\epsilon_k\approx \cos(k-ig)$ (recall $t = 1$) and are distributed on an ellipse
\begin{equation}
  \left(\frac{\Re(\epsilon)}{\cosh g}\right)^2 + \left(\frac{\Im(\epsilon)}{\sinh g}\right)^2 = 1
\end{equation}
(cf Fig.~\ref{fig:spec1}).

For the Hamiltonian \eqref{eq:ham} in the single-particle sector,
a heuristic relationship between the critical tilt $g_c$ and the end-to-end Green's function of an open chain was established in Ref.~\onlinecite{shnerb_winding_1998}.
There it was shown that the critical tilt of a ring is
\begin{equation}
\left(e^{N g_c}+e^{-Ng_c} -2\right)^{-1}= \frac{\prod\limits_{i=1}^{N} t_i}{\prod\limits_{i=1}^{N} (E-\epsilon_i)}
\end{equation}
with $\epsilon_i$ the spectrum of the ring (in the absence of a tilt), $E$ the energy at which the first eigenvalue develops an imaginary part, and $t_i$ are the hoppings between site $i$ and $i+1$ modulo $N$.
The right-hand side of this relationship is suggestive: were the $\epsilon_i$ eigenvalues of the \emph{open} chain at $g = 0$ and $g = g_c$ respectively, it would be closely related to the end-to-end Green's function of that open chain at energy $E$.
Since the eigenvalues $\epsilon_i$ will, in fact, approach the eigenvalues of the open chain in the long-system, strong-disorder limit, this provides good intuition---but the connection is definitely not exact.

\begin{figure}[h]
  \centering
  {\hskip -1.3cm
    \includegraphics[width=0.35\textwidth]{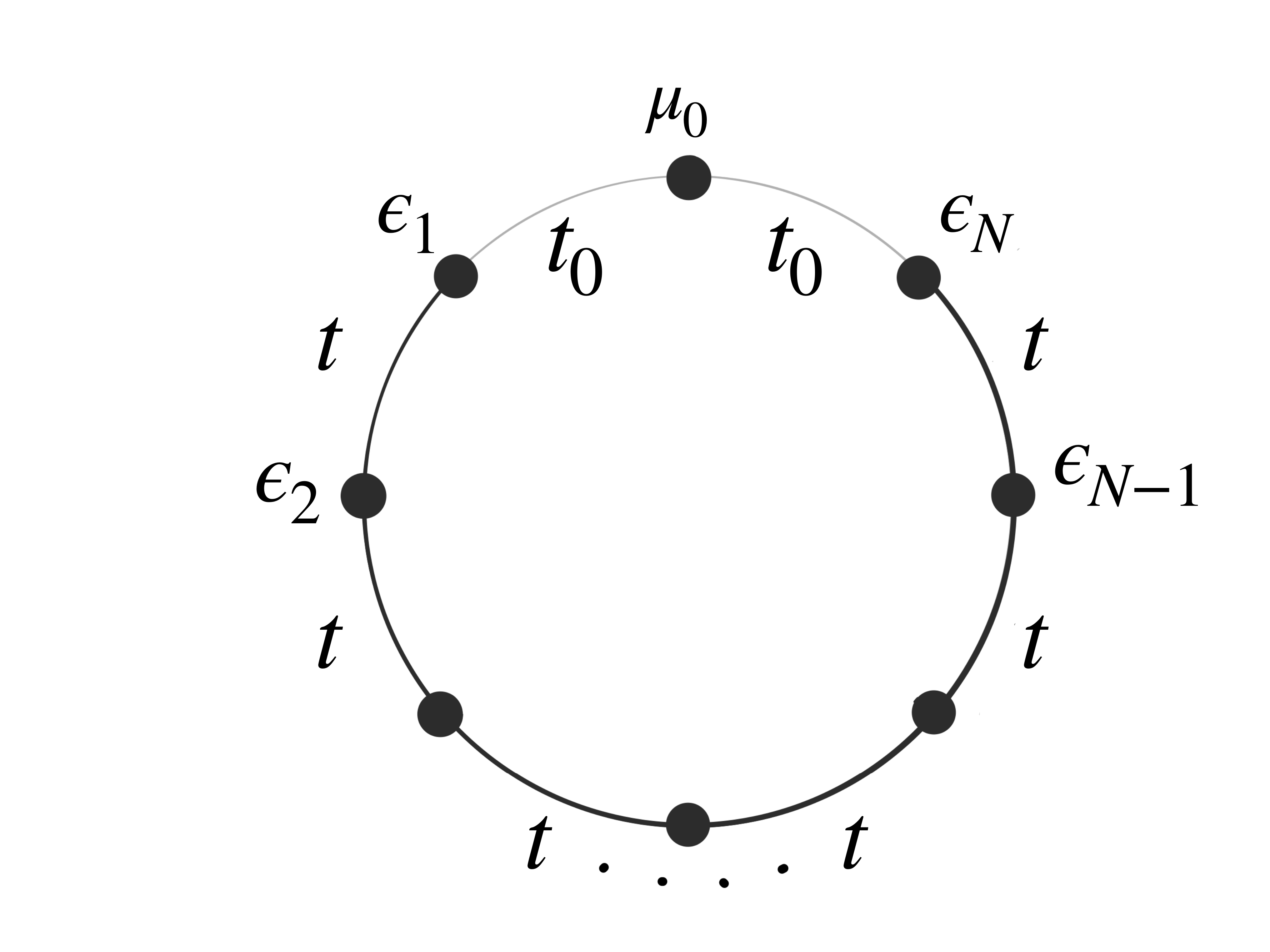}
  }
  \caption{\textbf{Lead setup for Sec.~\ref{s:sp}}: open chain (sites 1 through $N$ with onsite chemical onsite potentials $\epsilon_1 \dots \epsilon_N$ and uniform hopping amplitude $t$) together with lead site (chemical potential $\mu_0$) connected to chain by a tunneling $t_0$.}
  \label{fig:chain-with-lead}
\end{figure} 

A precise relationship between the critical tilt of a ring and the inverse localization length of  an open chain does exist;
we work it out in this section.
To expose the relationship we start with an open chain, and add a ``lead'' site with some local potential $\mu_0$ (cf. Fig.~\ref{fig:chain-with-lead}). The lead site is then connected weakly (with hopping $t_0\ll 1$) to both the first and the last sites of the open chain. Next, we calculate the determinant of the resulting closed chain in terms of the open-chain eigenvalues.
We then connect this determinant on the one hand to the end-to-end Green's function of the open chain, and hence to end-to-end eigenstate correlations; and on the other to the critical tilt.
We ultimately find that for an appropriate choice of chemical potential $\mu_0$ and tunneling strength $t_0$,
\begin{equation}
  e^{-(N+1) g_c} \sim \bra n c^\dagger_N c_1 \ket{n}
\end{equation}
where $\ket n$ is an eigenstate selected by $\mu_0$ of the open-chain Hamiltonian, while $1$ and $N$ are basis states on the first and last sites of the open chain.

Having established this relationship, we go on to generalize to generic lattice rings and to the many-particle (but noninteracting: $U = 0$) case.

\subsection{Determinant formula for the closed chain with lead}\label{s:sp:det}
Start with the Hamiltonian \eqref{eq:ham} in the $g = 0,\,U = 0$, open boundary conditions case---call it
\begin{align}
  \begin{split}
    \label{eq:openham}
    \Hopen^{[1:N]} &= t \sum_{j = 1}^{N-1} [c^\dagger_j c_{j + 1} + c^\dagger_{j+1}c_j ] + \sum_{j = 1}^{N} h_j n_j\;.\\
  \end{split}
\end{align}
(We write
$\Hopen^{[k:l]} = t \sum_{j = k}^{l-1} [c^\dagger_j c_{j + 1} + c^\dagger_{j+1}c_j + n_j n_{j+1}] + \sum_{j = k}^{l} h_j n_j$
for the Hamiltonian on sites $k$ through $l$ with open boundary conditions;
we will have occasion to use not only $\Hopen^{[1:N]}$ but also $\Hopen^{[2:N]}$.)

Add a ``lead'' site with chemical potential $\mu_0$ connected to both ends of the chain by a hopping amplitude $t_0$:
\begin{align} \begin{split}
    \label{eq:leadham}
    H = \Hopen^{[1:N]} + \mu_0 n_0
    + &t_0 (c^\dagger_0 c_1 + c^\dagger_1 c_0)\\
    + &t_0 (e^{(N+1)g}c^\dagger_N c_0 + e^{-(N+1)g}c^\dagger_0 c_N)\;.
  \end{split} \end{align}
Since the chain now has periodic boundary conditions,
we can no longer gauge away the imaginary vector potential \`{a} la \eqref{similarity};
it is convenient to work in a gauge in which all of the vector potential lives on the bond between the lead and site $N$.
We ultimately plan to take $N \gg 1$ and $g \gtrsim 1$, so we can comfortably ignore the term $t_0 e^{-(N+1)g}c^\dagger_0 c_N$.

For the purposes of finding a precise determinant formula,
we take the lead to be \emph{weakly} connected to the rest of the chain: $t_0 \ll t$.
We will discuss relaxing this assumption below.

$H$ then has matrix representation
\begin{align}
  \begin{split}
    &EI - H(g)\\
    &\ =
    \begin{bmatrix}
      E - \mu _0 & t_0 & & &  \\
      t_0 & E - \epsilon _1 & t &   \\
      & t & E - \epsilon _2 & t &   \\
      && t & E - \epsilon _3 & t &  \\
      t_0e^{(N+1)g}&&& t & E - \epsilon _4 &  \\
    \end{bmatrix}
  \end{split}
\end{align}
and determinant
\begin{align}
  \label{eq:lead-det}
  \begin{split}
    \det(EI -  H(g)) &= (E - \mu _0) \det (EI - \Hopen)\\
    &\quad\quad - t_0^2 \det (EI - \Hopen^{[2:N]})\\
    &\quad\quad + (-1)^{N+2} t_0^2 t^{N-1}e^{(N+1)g}
  \end{split}
\end{align}
(expanding in minors along the first column). 
Since we take $t_0$ small we can ignore the $t_0^2$ term compared to the $t_0^2 e^{gN}$ term.
If we take $E$ to be an eigenvalue of $H(g)$ this is 
\begin{align}
  \label{eq:det-pbc}
  \begin{split}
    &\det (EI - \Hopen^{[1:N]})\\
    &\quad =  (-1)^N t^{N-1} (E - \mu _0)^{-1} t_0^2 e^{(N+1)g}\;.
    \end{split}
\end{align}

\subsection{Open-chain Green's function}\label{s:sp:greens}
We can re-write the determinant in \eqref{eq:det-pbc} in terms of the open-chain Green's function.
This has $(1,N)$ matrix element
\begin{align}
  \begin{split}
    G^{[1:N]}_{1N}(E) &= [(EI - \Hopen^{[1:N]})^{-1}]_{1N}\\
    &= \frac{[\adj(EI - \Hopen^{[1:N]})]_{1N}}{\det (EI - \Hopen^{[1:N]})}
  \end{split}
\end{align}
for 
\begin{equation}
  \label{eq:green-adj}
  G^{[1:N]}_{1N}(E) = (-1)^{N+1}[\det (EI - \Hopen^{[1:N]})]^{-1}  t^{N-1}\;.
\end{equation}
With this relation \eqref{eq:det-pbc} becomes
\begin{equation}
  \label{eq:gns-expg}
  (G^{(1:N]}_{1N})^{-1} = -(E - \mu _0)^{-1} t_0^2 e^{(N+1)g}\;.
\end{equation}
We can extract the tunneling probability for an eigenstate $\ket \alpha$ of $\Hopen^{[1:N]}$ from the Green's function 
\begin{equation}
  \braket{1|\alpha}\braket{\alpha|N} = (E - E_\alpha) G_{1N}^{[1:N]}(E)
\end{equation}
by identifying poles,
so Eq.~\eqref{eq:gns-expg} is
\begin{equation}
  \label{eq:shnerb-like-ampl}
  e^{-(N+1) g} = \langle 1 | \alpha\rangle \langle \alpha | N \rangle ( E_\alpha - E)^{-1} (E - \mu _0)^{-1} t_0^2 \;.
\end{equation}

\subsection{Critical tilt $g_c$}\label{s:sp:gc}

\begin{figure}[t]
  \centering
  \includegraphics[width=0.45\textwidth]{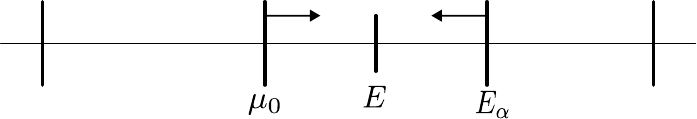}
  \caption{\textbf{Sketch of the lead spectrum} as in Sec.~\ref{s:sp:gc}: $E_\alpha$ is a level of the open chain, $\mu_0$ the chemical potential of the lead, and $E$ the energy at which we measure the Green's function. Unlabeled vertical lines are other eigenenergies of the open chain.}
  \label{fig:lead-eigenvalues}
\end{figure} 

Eq.~\ref{eq:shnerb-like-ampl} has three free parameters: $\mu_0$, $t_0$, and $g$.
$E$ is not a (continuously tunable) parameter:
it is fixed by $\mu_0, t_0, g$,
since it is an eigenvalue of the non-Hermitian Hamiltonian with lead site. 
But we can choose these parameters to strongly constrain the non-Hermitian eigenvalue $E$,
and hence relate $g_c$,
the tilt at which eigenstate $\alpha$ coalesces with the lead state and develops an imaginary part,
to $\langle 1 | \alpha\rangle \langle \alpha | N \rangle$.

Suppose we wish to probe the eigenstate $\alpha$ of $\Hopen^{[1:N]}$. Then choose
\begin{align}
  \label{eq:lead-params}
  \begin{split}
  0 < &E_\alpha - \mu_0  \ll \text{typical level spacing}\;,\\
  t_0 &= \frac 1 2 (E_\alpha - \mu_0)\;
  \end{split}
\end{align}
(cf Fig.~\ref{fig:lead-eigenvalues}).
Because the open chain is localized, the lead's occupied state will not hybridize substantially with any of the chain's levels in the Hermitian chain.
But as we increase $g$, the lead state and the chain level $n$ will start to hybridize, and the energy of the lead site and of state $n$ will approach each other.
When they coalesce, which they will do at a value $E \simeq \frac 1 2 (\mu_0 + E_n)$, both levels will develop imaginary parts.
Because $ E_\alpha - \mu_0\ll \text{typical level spacing}$, we expect this to be the first pair to coalesce.
With
\begin{equation}
  \label{eq:limit-rail}
  \gamma \equiv \frac {t_0^2}{(E_\alpha - E)(E - \mu_0)} \simeq 1
\end{equation}
(where the estimate follows from our premeditated choice of $t_0$),
Eq.~\ref{eq:shnerb-like-ampl} will become
\begin{align} \begin{split}
    \label{eq:gc-xi}
  g_c &= \frac{1}{N+1} \ln \left[\gamma\bra \alpha c^\dagger_1 c_N\ket \alpha\right]\\
    &= \xi_\alpha^{-1} + \frac {\ln \gamma}{N+1} \simeq \xi_\alpha^{-1}
  \end{split} \end{align}
where we define an eigenstate localization length $\xi_n^{-1} \equiv \ln\bra \alpha c^\dagger_1 c_N\ket \alpha$.

\begin{figure}[t]
  \centering
  \includegraphics[width=0.45\textwidth]{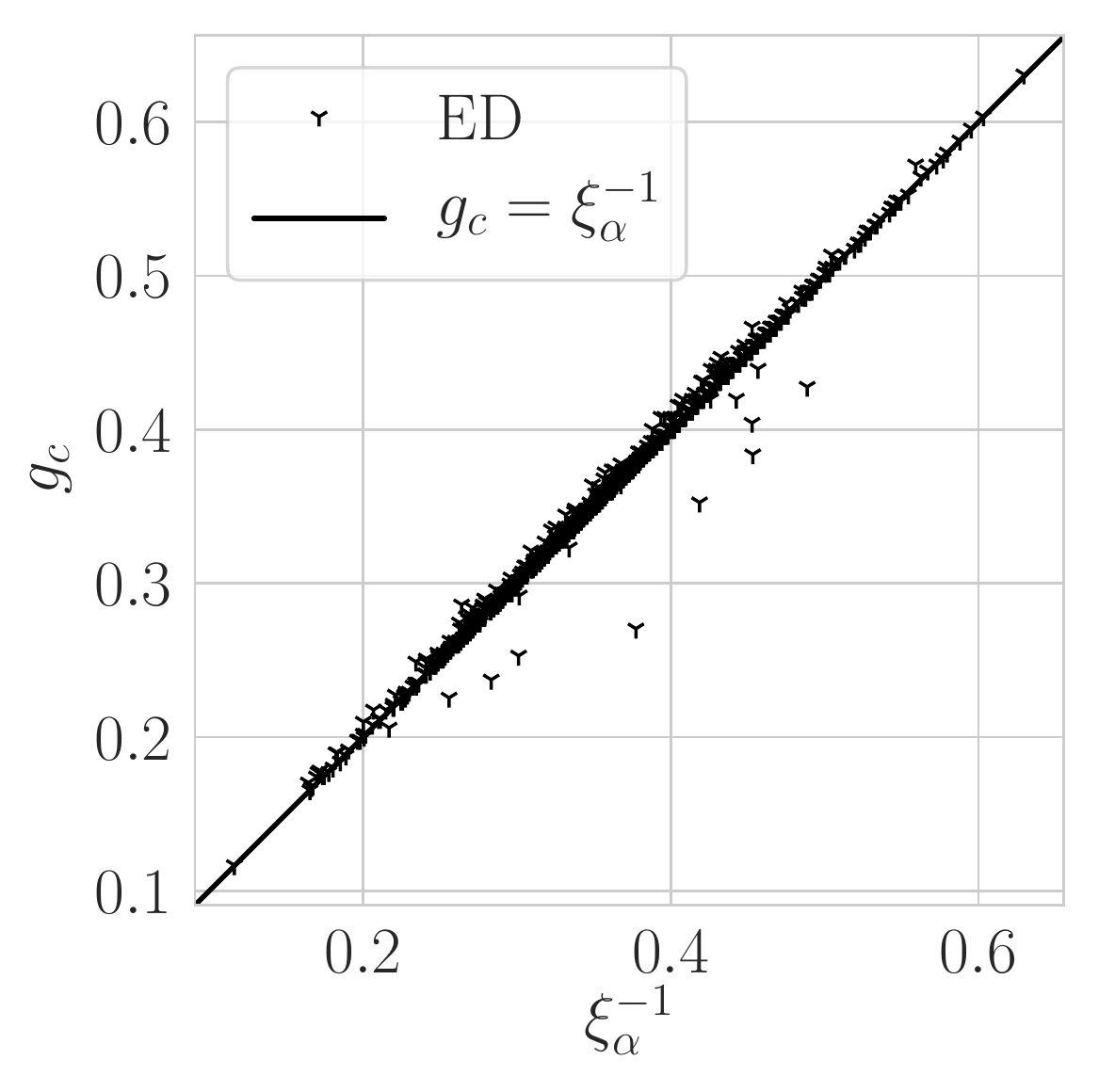}
  \caption{\textbf{Critical tilt $g_c$  against inverse localization length $\xi_\alpha^{-1}$} via exact diagonalization for 1000 disorder realizations of \eqref{eq:leadham} at system size $N = 40$ and lead parameters as in Eq.~\eqref{eq:lead-params}. (We consider eigenstate $\alpha = 20$ of each realization.) This confirms our analytical relationship \eqref{eq:gc-xi}.}
  \label{fig:gc-xialpha-lead}
\end{figure}

We show $g_c$ and $\xi_\alpha$ for eigenstate $\alpha = 20$ of a chain with $N=40$ with 1000 disorder realizations in Fig.~\ref{fig:gc-xialpha-lead},
and see good agreement.
We first diagonalize the open chain;
we then take $\mu_0 = E_{20} - 0.01 h/L$ and $t = 0.005 h/L$, in accordance with Eq.~\eqref{eq:lead-params},
and find $g_c$ in the resulting closed chain.
The variation comes from $\gamma$: $E$ is not always exactly $E = \frac 1 2 (E_\alpha + \mu_0)$.

\subsection{Closed chain without lead}\label{s:sp:wol}
Even though the chain with lead has periodic boundary conditions,
in the sense that there are exactly two paths between any two sites,
it is not obvious that the results of Sec.~\ref{s:sp:gc} will carry over to ordinary chains with periodic boundary conditions.
Eq.~\ref{eq:gc-xi}, which connects $g_c$ and $\xi_\alpha$ for some eigenstate,
requires a carefully fine-tuned lead site.
Can we do better? Can we take a generic lead site---that is, a straightforward periodic chain?

If we are willing to relax our demands for rigor,
we can make some estimates.
Take the Hamiltonian \eqref{eq:ham} with periodic boundary conditions acting on one particle.
Single out one arbitrary site for treatment as the ``lead'', and return to \eqref{eq:lead-det}.
Once again take $E$ to be an eigenvalue of the non-Hermitian Hamiltonian $H(g)$,
so \eqref{eq:lead-det} becomes
\begin{align}
  \begin{split}
    0  &= (E - \mu _0) \det (EI - \Hopen^{[1:N]}) - t_0^2 \det (EI - \Hopen^{[2:N]})\\
    &\quad\quad + (-1)^{N+1} t_0^2 t^{N-1}e^{(N+1)g}\;.
  \end{split}
\end{align}
Take $t_0 = t$---the supposed lead site is just a normal lattice site, after all---and write the determinants in terms of the $(1,N)$ components of the Green's functions $G^{[1:N]}, G^{[2:N]}$ of $\Hopen^{[1:N]}, \Hopen^{[2:N]}$.
This becomes
\begin{align}
  \begin{split}
    0 &= (E - \mu_0) \braket{1|G^{[1:N]}|N}^{-1} + t\braket{2|G^{[2:N]}|N}^{-1}\\
    &\quad +  e^{(N+1) g}\;.
  \end{split}
\end{align}
Now work at $g_c$.
Once again write $E_\alpha$ for the eigenvalue nearest $\mu_0$;
even though $t_0$ is no longer small, we expect
\begin{equation}
  E - \mu_0 \sim E_\alpha - E \sim \frac 1 2 (E_\alpha - \mu) \sim \frac 1 {\sqrt{L}} \ll 1 \;,
\end{equation}
so we can ignore the $G^{[1:N]}$ term.
If we assume
\begin{equation}
  \label{eq:2N-gns-fn}
  \braket{2|G^{[2:N]}|N}^{-1} \sim e^{-(N - 1) \xi^{-1}}\;
\end{equation}
then
\begin{equation}
  g_c \sim \xi_\alpha^{-1}\;.
\end{equation}

\section{Many-particle case and interaction broadening}\label{s:mp}
\subsection{Many-particle non-interacting case}\label{s:mp:nonint}

\begin{figure}[t]
	\raggedleft
	\includegraphics[width=0.45\textwidth]{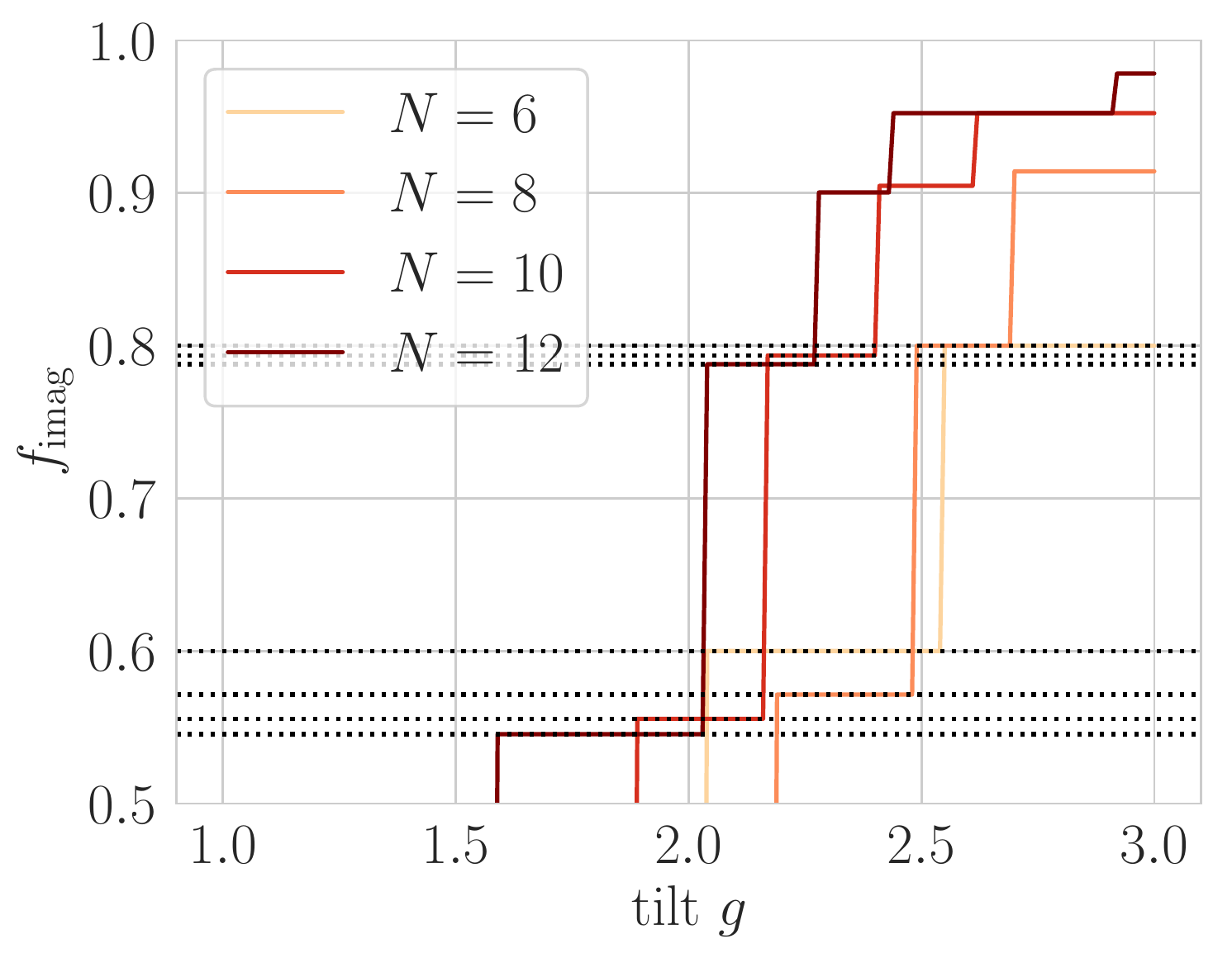}
	\caption{
          \textbf{Emergence of imaginary part of energy with increasing tilt $g$.}
          $y$ axis is fraction of eigenvalues having imaginary component for one realization of \eqref{eq:ham} at interaction strength $U = 0$, disorder width  $W = 7$, and system size $N$ as indicated.
          Dotted lines show our analytical results \eqref{eq:sp-nonint-num1}, \eqref{eq:sp-nonint-num2} for the first few bifurcations.
          }
	\label{fig:imagsteps} 
\end{figure} 
Now let the same Hamiltonian \eqref{eq:ham} act on many particles---in fact on the half-filling sector---but take it to be noninteracting ($U = 0$).
Its eigenstates will be Slater determinants $\prod_{\alpha \in A} c^\dagger_\alpha \ket 0$ with eigenvalues $E_A = \sum_{\alpha \in A} E_\alpha$.
When two single-particle states pass through an exceptional point, developing imaginary parts to their energies, they therefore take with them a whole class of many-particle states.

To quantify this effect consider first increasing $g$ through $g_c$, the tilt at which the first two single-particle states go through an exceptional point.
(In the example above, of an open chain with a lead site, these will be the lead site and the open-chain level $n$.)
Call those two states $\alpha_1$ and $\alpha_2$, and occupy a set {\it A} of additional levels, not including $\alpha_1, \alpha_2$, with more particles. 
Since the energy difference of the many body state is the same as that of the delocalizing orbitals, 
\begin{equation}
  \label{eq:many-nonint-deltaE}
  E_{\alpha_1  A}(g) - E_{\alpha_2  A} (g)= E_{\alpha_1}(g) - E_{\alpha_2}(g)
\end{equation}
\emph{every} such set gives a pair of levels that coalesce at $g_c$.
As we tune $g$ through $g_c$, then, all
\begin{equation}
  \label{eq:sp-nonint-num1}
  n_1 =2\cdot {N - 2 \choose N/2-1}
\end{equation}
levels with either $\alpha_1$ or $\alpha_2$ occupied will coalesce with the states with $\alpha_2$ and $\alpha_1$ occupation switched.
(Recall that we assume a half filled system with an even number $N$ of sites.)
These states will re-emerge with imaginary parts, simply because the energies are the sum of the single-particle energies $E_{\alpha}$.
(Note that if both $\alpha_1$ and $\alpha_2$ are occupied the resulting energy is real, because $E_{\alpha_2} = E^*_{\alpha_2}$.)

Consider now increasing $g$ through the tilt at which the second pair of single-particle eigenstates passes through an exceptional point.
At this tilt
\begin{equation}
  \label{eq:sp-nonint-num2}
 n_2= 2 \cdot {N - 4 \choose N/2 - 1}+2\cdot {N - 4 \choose N/2 - 3}=4\cdot {N - 4 \choose N/2 - 1}
\end{equation}
eigenstates will develop imaginary parts (these two expressions have $\alpha_1$ and $\alpha_2$ either fully occupied or fully unoccupied).

Fig. \ref{fig:imagsteps} shows the fraction of eigenenergies that develop a complex eigenenergy as a function of disorder for a particular disorder and no interactions.

\subsection{Many-particle interacting case}\label{s:mp:int}

\begin{figure}[t]
	\raggedleft
	\includegraphics[width=0.45\textwidth]{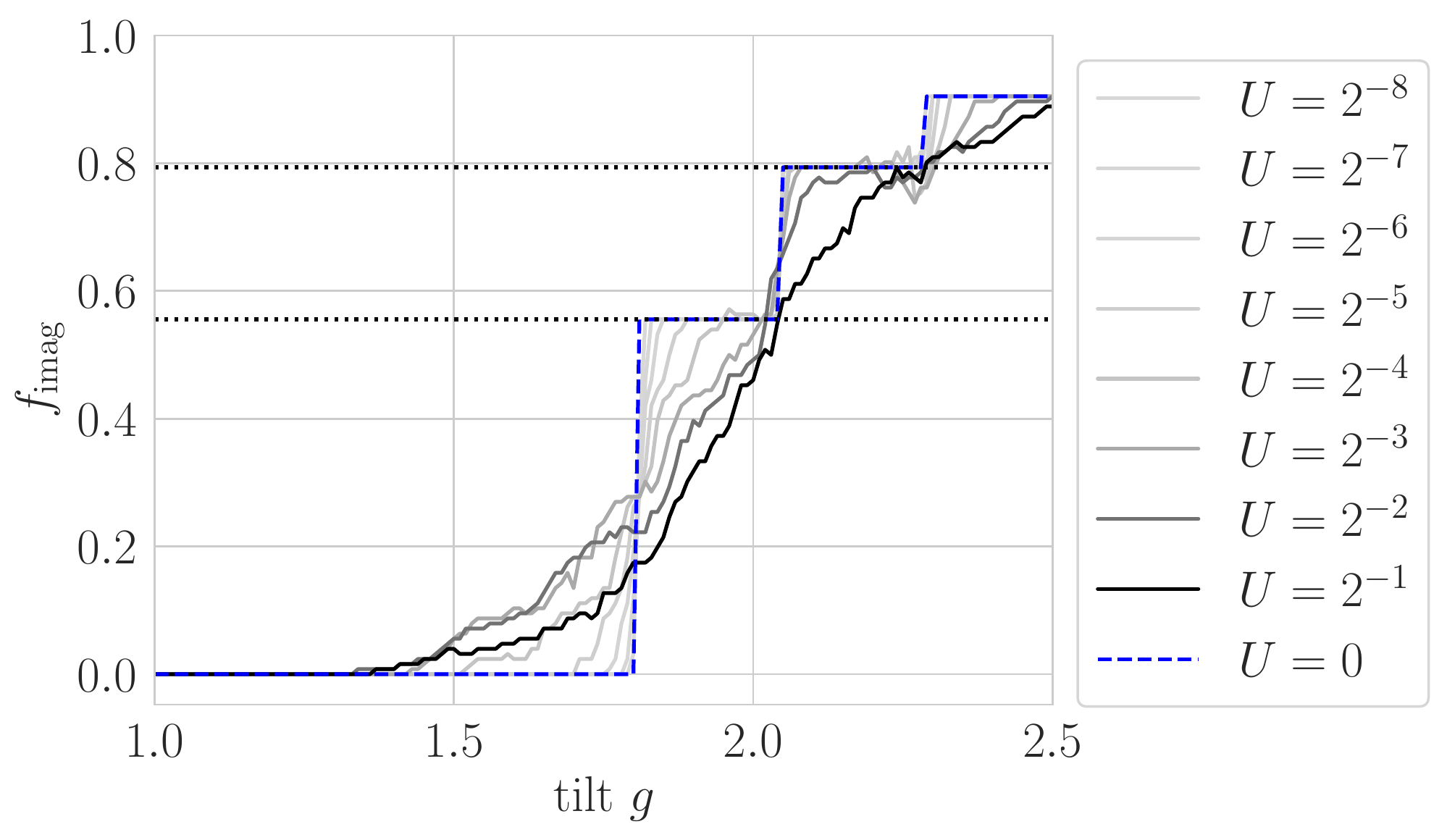}
	\caption{
          \textbf{Effect of weak interaction on emergence of imaginary parts of energy}. $y$ axis is fraction of eigenvalues having imaginary component for one realization of \eqref{eq:ham} at disorder width $W = 8$, system size $N = 10$, and interaction strength $U$ as indicated.
          Dotted lines again show the analytical results \eqref{eq:sp-nonint-num1}, \eqref{eq:sp-nonint-num2}.
          In the MBL phase ($U = 1,\,W = 8$) interactions smear out the discrete steps characteristic of the noninteracting case, which result from coalescence of single-particle eigenstates.
          }
	\label{fig:imagstepsint} 
\end{figure}

Turn now to the interacting case, and consider disorder strong enough that the  Hamiltonian \eqref{eq:ham} is fully localized for $0 \le U \le 1,\,g = 0$.
In terms of $\ell$-bits that interacting Hamiltonian is 
\begin{align}
  \label{eq:lbit}
  H = \sum E_j \tilde n_j
  &+ \sum_{jk} J^{(2)}\tilde n_j\tilde n_k \\
  &+ \sum_{j_1 j_2 j_3} J^{(3)}_{j_1j_2j_3}\tilde n_{j_1}\tilde n_{j_2} \tilde n_{j_3} + \cdots  \nonumber
\end{align}
In the single-particle sector this reduces to $H(g) = \sum_\alpha E_\alpha n_{\alpha}(g)$.

One can imagine running the same procedure as in the previous part.
As one increases $g$, the single-particle eigenvalues develop imaginary parts---but this cannot lead to simultaneous coalescence of many eigenvalues.
The interaction terms $\sum_{\alpha\beta} J^{(2)}_{\alpha\beta} \tilde n_\alpha \tilde n_\beta + \dots$ mean that now
\begin{equation}
  \label{eq:many-int-deltaE}
  E_{\alpha_1  A}(g) - E_{\alpha_2  A} (g)\ne E_{\alpha_1} - E_{\alpha_2}\;,
\end{equation}
in contrast to \eqref{eq:many-nonint-deltaE},
in which the energy difference was independent of the additional orbitals $A$.
The $\ell$-bit interactions of Eq.~\eqref{eq:lbit} therefore smooth the sharp step-like coalescence of many body states;
the degree of this smoothing probes the strength of those interactions.

\section{Phase diagram of the random-field XXZ model}\label{s:phase}
In this section we make use of the relationship $g_c \sim \xi^{-1}$ to probe the phase diagram of the model \eqref{eq:ham} using the critical tilt.
We show that it is consistent with previous studies.

\subsection{Fixed interaction}\label{s:phase:iso} 

\begin{figure*} [t]
  \begin{minipage}{0.45\textwidth} 
    \includegraphics[width=\textwidth]{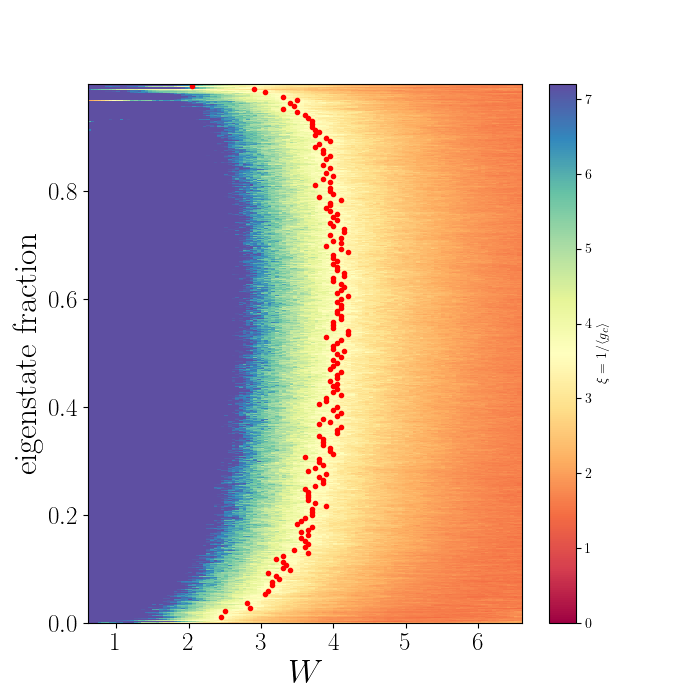}
  \end{minipage}
  \begin{minipage}{0.45\textwidth} 
    \includegraphics[width=\textwidth]{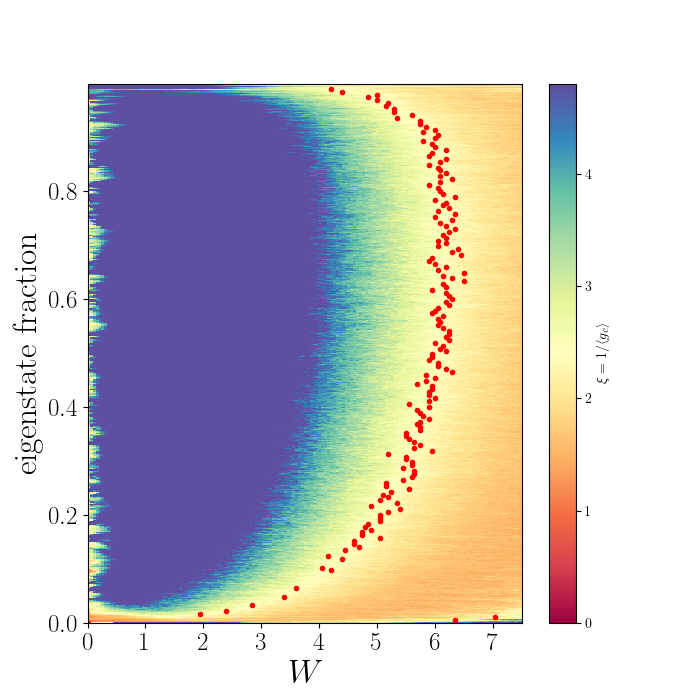}
  \end{minipage}
  
    \caption{
      \textbf{Phase diagram of the disordered, interacting Hamiltonian \eqref{eq:ham} }
      at
      system size $N = 12$,
      interaction strength $U = 1$ (\textbf{left}) and $U = 3$ (\textbf{right}),
      half filling,
      extracted from critical tilts.
      The color scale is $\xi \equiv [\bar g_c]^{-1}$;
      we show it as a function of disorder width $W$
      and eigenstate fraction with eigenstates sorted from lowest to highest.
      See Fig.~\ref{fig:phd6o12-e} for localization length as a function of energy.
      Red dots indicate $\xi = 0.3N$ ($U = 1$, left) or $\xi = 0.2N$ ($U = 3$, right),
      which is consistent with the crossing in the scaling collapse of Fig. \ref{fig:scale}.
      Compare to \onlinecite{PhysRevB.91.081103} Fig.~1;
      see Fig.~\ref{fig:phd6o12-e} for the same data plotted against energy.
    }
    \label{fig:phd6o12} 
\end{figure*}

Considering the critical tilt $g_c$ for each eigenstate gives us the localization length as a function of energy.
We measure $g_{c;r\alpha}$ for each eigenstate $j$ of each disorder realization $r$ with precision $0.05$; we then average before inverting to estimate a localization length:
\begin{align}
  \begin{split}
    \xi_j &:= [\bar g_c]^{-1}\\
    &= \left[\frac 1 {N_{\mathrm{realizations} } } \sum_{\text{realizations $r$} } g_{c;rj} \right]^{-1}\;.
  \end{split}
\end{align}
In Fig.~\ref{fig:phd6o12} we show $\xi_j$ as a function of eigenstate fraction $j {N \choose N/2}^{-1}$.
We mark 
\begin{equation}
  \label{eq:phase-criterion}
  \xi_j = c L
\end{equation}
with $c$ chosen via finite-size scaling;
this gives a heuristic estimate of the phase transition.

The resulting phase diagram is broadly consistent with that of \onlinecite{PhysRevB.91.081103}.
We see an apparent mobility edge for $1 \lesssim W \lesssim 4$, as well as full localization (per our criterion \eqref{eq:phase-criterion}) for $W_c \simeq 4$.
(Our critical disorder is different because we work at a different interaction strength.)
We also see a slight asymmetry in $\Re E \leftrightarrow - \Re E$, again consistent with \onlinecite{PhysRevB.91.081103}.

\begin{figure}[t]
  \begin{minipage}{0.45\textwidth}
    \includegraphics[width=\textwidth]{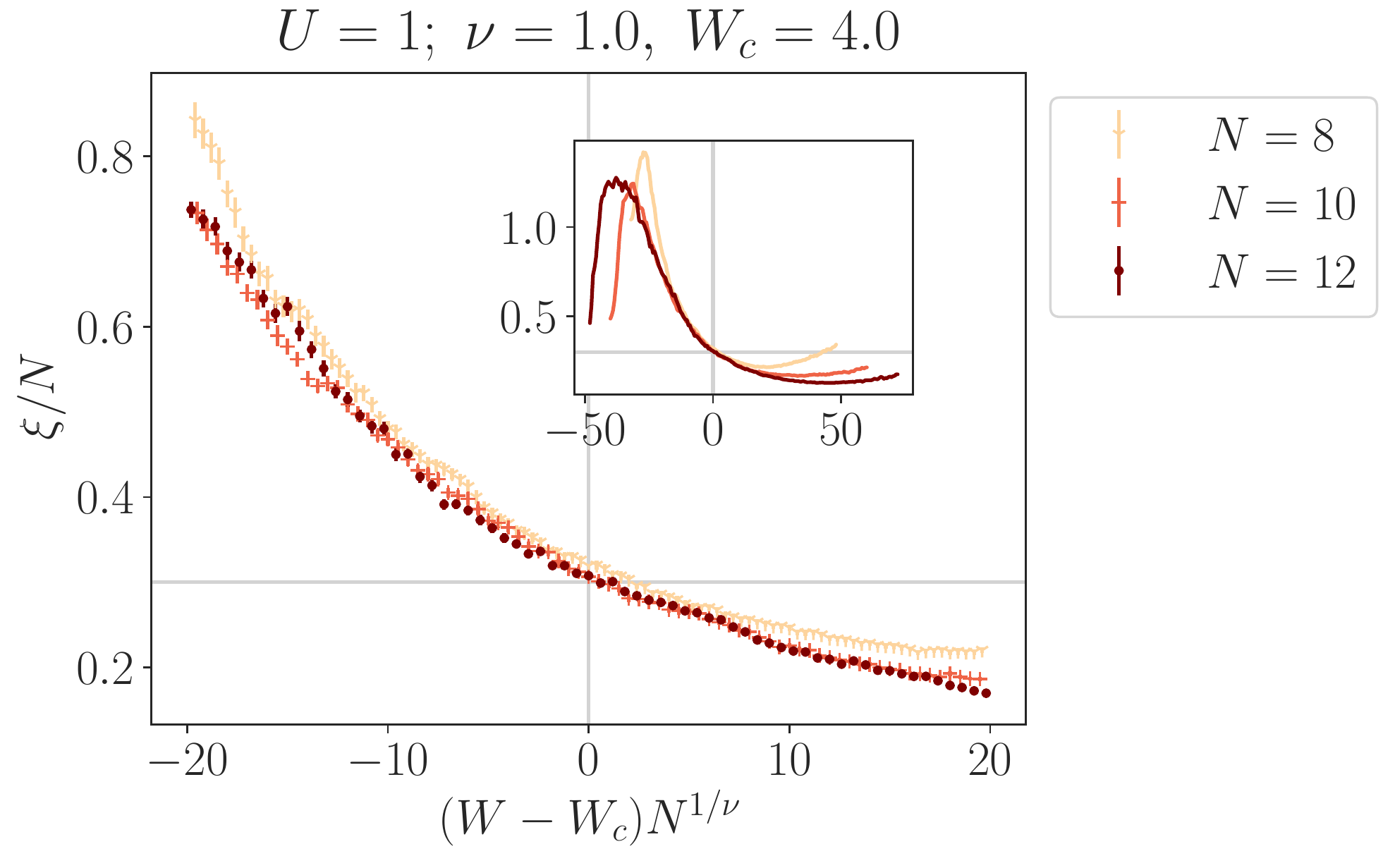}
  \end{minipage}
  \begin{minipage}{0.45\textwidth}
    \includegraphics[width=\textwidth]{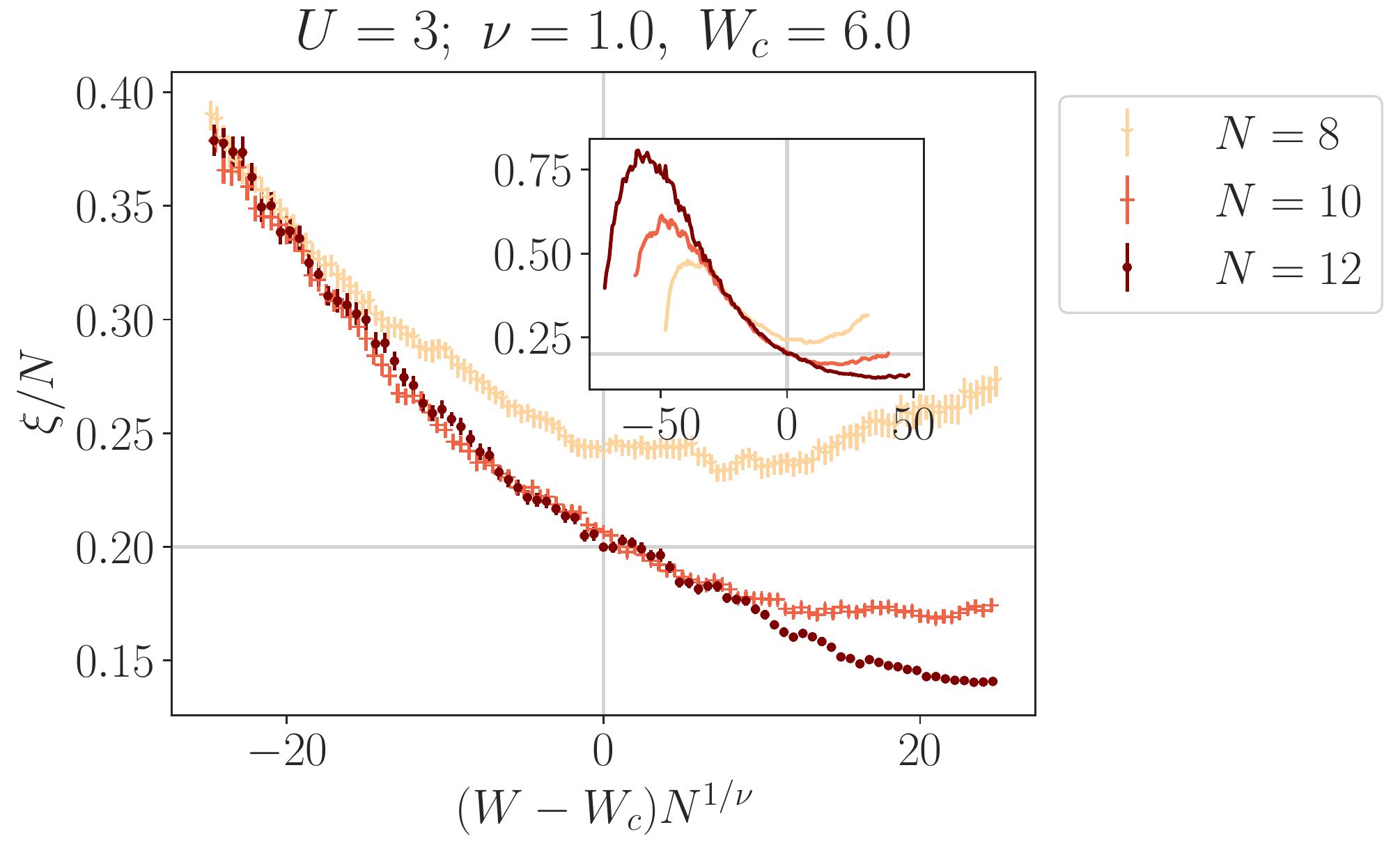}
  \end{minipage}
        
    \caption{
      \textbf{Finite-size scaling collapse}
      for the localization length $\xi = [\bar g_c]^{-1}$
      at interaction strength $U = 1$ (\textbf{top})
      and $U = 3$ (\textbf{bottom})
      with critical disorder width $W_c = 4$ and correlation-length exponent $\nu = 1$.
      We average over 10 eigenstates near the middle of the spectrum.
      We find a crossing at $\xi = 0.3 N$ and $\xi = 0.2 N$ (where $N$ is system size)
      for $U = 1$ and $U = 3$, respectively.
      The grey horizontal lines indicate those crossings.
      We show putative scaling collapse for different $W_c$, $\nu$ in Appendix \ref{a:scale}.
      These scalings result from the average of 99 ($U = 1$) or 98 ($U = 3$) disorder realizations.
      Errorbars come from nonparametric bootstrap.
    }
    \label{fig:scale}
  \end{figure}

Finite-size scaling gives a better estimate for $W_c$, as well as an estimate for the correlation length exponent $\nu$.
In addition to averaging over disorder realizations, we average over 10 eigenstates $n_0 = 0.5{N \choose N/2}$ through $ n_0+9 = 0.5{N \choose N/2 } + 9$ near the middle of the spectrum: 
\begin{align}
  \begin{split}
    \xi &= [\bar g_c]^{-1}\\
    &= \left[\frac 1 {10 N_{\mathrm{realizations} } } \sum_{\text{realizations $r$} } \sum_{j = n_0}^{n_0 + 9} g_{c;rj} \right]^{-1}\;.
  \end{split}
\end{align}
This gives cleaner statistics, but does not appreciably change the scaling parameters we extract.
By seeking a scaling collapse (Fig.~\ref{fig:scale}---cf App.~\ref{a:scale}),
we find $W_c \approx 4$ and $\nu \approx 1$.
Our system sizes are very small,
so we do not claim this scaling collapse reflects the ultimate large-system properties of the transition.
(In looking for ultimate large-system behavior, we would need to in addition check for Kosterlitz-Thouless behavior. \cite{dumitrescu_scaling_2017,morningstar_renormalization-group_2019,dumitrescu_kosterlitz-thouless_2019,goremykina_analytically_2019,suntajs_ergodicity_2020})
Nevertheless, even at these small sizes our collapse is not consistent with the result of Hamazaki et al.,
who find a correlation-length exponent $\nu = \frac 1 2$.\cite{hamazaki_non-hermitian_2019}.

Like us, Ref. \onlinecite{hamazaki_non-hermitian_2019}, Hamazaki et al., investigates a PT-breaking transition in a localized many-body system, with a finite non-Hermitian tilt. Our measurements, however, differ from those of Ref. \onlinecite{hamazaki_non-hermitian_2019} both ontologically and operationally.
Ontologically, Hamazki et al., treat the non-Hermitian Hamiltonians as objects of study in their own right.
They fix $g \ne 0$ and look for a phase transition as a function of disorder width, $W$.
We, by contrast, use non-Hermitian Hamiltonians as indicators of the properties of the underlying $g = 0$ Hermitian Hamiltonian. Particularly, we are seeking to explore the properties of the delocalization transition of the $g=0$ system, and our scaling plots refer to the $g=0$ transition only. The $g=0$ transition  could well have different universal properties than the $g>0$ disorder-tuned transition that Ref. \onlinecite{hamazaki_non-hermitian_2019}  is studying.  Less importantly, operationally, Ref. \onlinecite{hamazaki_non-hermitian_2019}  measures the fraction of eigenvalues with imaginary parts, whereas we measure the critical tilt for \emph{each} eigenvalue and average.

  A straightforward interpretation of our finite-size scaling (Fig.~\ref{fig:scale})
  implies that our localization length $\xi = g_c^{-1}$ diverges at the transition.
  This is in striking contrast to the avalanche theory of the localization transition,
  which posits a finite typical localization length at the transition \cite{thiery_many-body_2018,thiery_microscopically_2017} 

  The reason for the contrast is that the localization length used as a parameter in the avalanche picture measures the decay of matrix elements;
  the avalanche results from the competition between that decay and Hilbert space growth.
  Our $g_c^{-1}$, in contrast,
  measures the competition directly:
  it is a quantity with dimensions of length
  measuring a competition between matrix elements of the end-to-end hopping $c^\dagger_1 c_N$
  and the many-body Hilbert space dimension, characterized by the gaps between eigenstates.
  In the language of \onlinecite{abanin_colloquium_2019} Sec.~IV A,
    our $g_c^{-1}$ is
    \begin{equation}
      \label{eq:abanin-coll}
      \xi = g_c^{-1} = [s - l_*]^{-1}
    \end{equation}
    where $s$ is the entropy density
    and $l_*$ is the localization length associated with operator matrix elements.
  To see this, recall that before an eigenvalue can develop an imaginary part,
  it must become degenerate with another eigenvalue.
  So if we imagine gauging all of the flux to one bond and adding that bond perturbatively,
  we find that the change in energy induced by the term $te^{gN} c^\dagger_1 c_N$
  must be comparable to the gap between
  the eigenstate in question and one of its neighbors.
  This is precisely our argument leading up to Eq.~\eqref{eq:intuitive-argument},
  with now $\ket n$ a many-body eigenstate
  and $\Delta E$ the gap between $\ket n$ and another (many-body, interacting) eigenstate nearby in energy.
    From the expression \eqref{eq:abanin-coll}
    it is clear that our localization length $\xi = g_c^{-1}$
    can diverge even when the localization length $l_*$ associated with operator matrix elements is finite,
    and that our $g_c^{-1}$ should diverge at the critical value of $l_*$
    predicted by
    either the straightforward logic of \onlinecite{abanin_colloquium_2019}
    or the more detailed logic of the avalanche picture.

  $g_c^{-1}$ also immediately measures coherent end-to-end transport in a finite segment of a chain.
  We showed this explicitly in a non-interacting chain,
  but even in an interacting chain we can see by rearranging Eq.~\eqref{eq:intuitive-argument} to
  \begin{equation}
     e^{N g_c} \sim \bra{n} c^\dagger_N c_1 \ket n \Delta E^{-1} 
  \end{equation}
   that $g_c$ measures the magnitude of something having the form of an end-to-end Green's function.
    (Note once again that here $\ket n$ and $\Delta E$ are eigenstates and gaps of the many-body interacting Hamiltonian).

    We expect that the origin of the critical divergence of $g_c^{-1}$ is best understood in the context of long-range resonant structures;\cite{khemani_critical_2017,herviou_multiscale_2019,villalonga_eigenstates_2020}
    it may provide a useful diagnostic of those resonances.

\begin{figure}[t!]
	\raggedleft
	\includegraphics[width=0.50\textwidth]{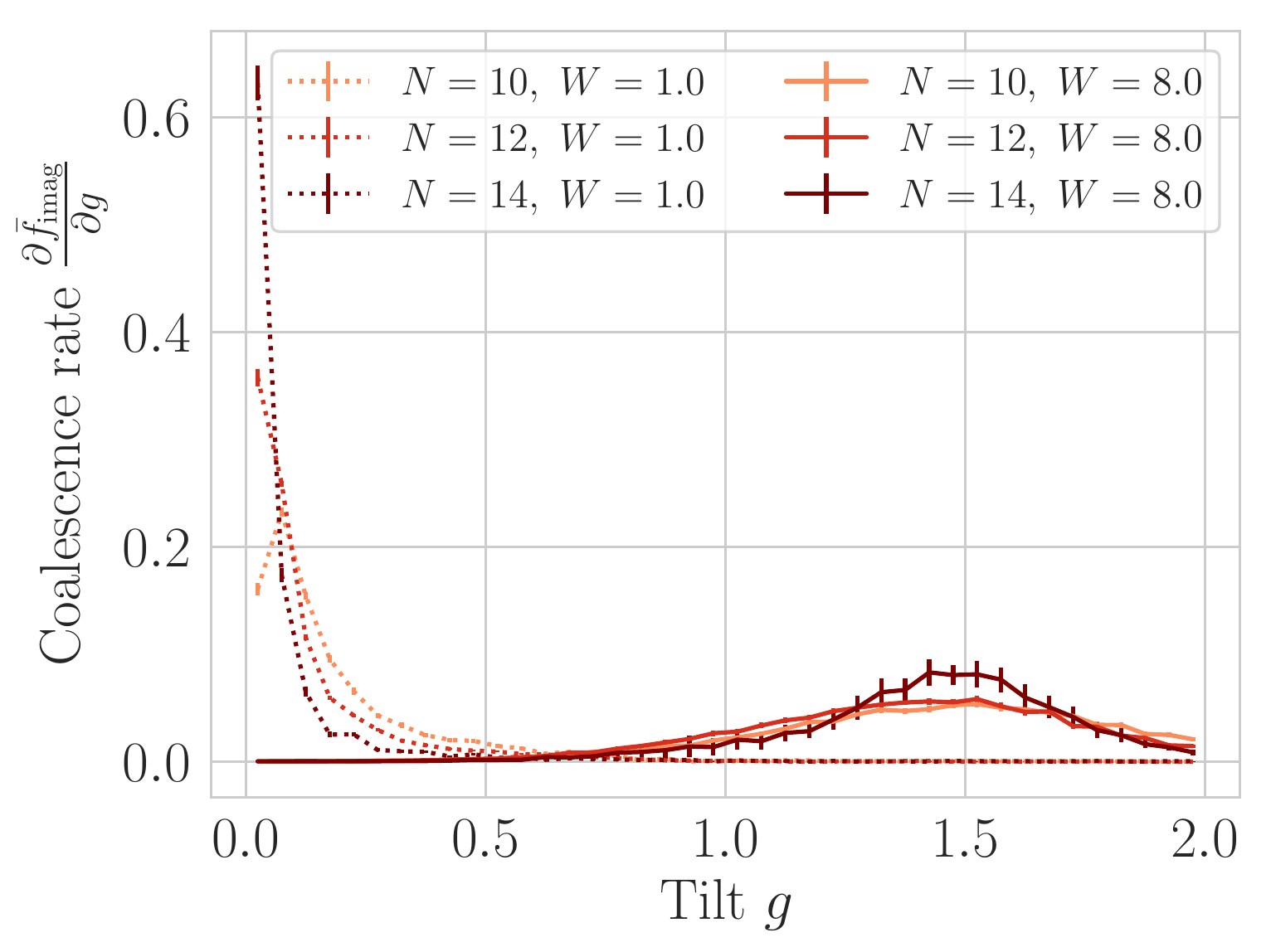}
	\caption{
          \textbf{Distribution of coalescence points}:
          rate of change $\frac{\partial}{\partial g} \bar f_\imag$
          of fraction of eigenenergies having imaginary parts
          as a function of $g$ for interaction strength $U = 1$ and disorder width $W = 1$ (dots), $W = 8$ (solid).
          The average is over $N_{\mathrm{realizations}} = 100$ disorder realizations for $N = 10,12$ sites
          and $N_{\mathrm{realizations}} = 10$ disorder realizations for $N = 14$ sites;
          errorbars are $\frac 1 {\sqrt{N_{\mathrm{realizations} }}}\mathrm{std}\; \frac {\partial f_{\mathrm{imag}}}{\partial g}$.
          In the ETH phase ($ W = 1$) the distribution is peaked near $g = 0$;
          in MBL phase ($W = 8$) it is peaked at some finite $g_{\max} \sim g_c$.
          This disorder average does not display a critical $g_c$,
          because for any $g$ there will be disorder realizations with critical $g_c < g$.
          The behavior of $ \left.\frac{\partial}{\partial g}\bar f_{\mathrm{imag}}\right|_{\mathrm{MBL}} $ near $g = 0$
          therefore indirectly probes rare ``quasi-thermal'' disorder realizations.
          This provides a diagnostic for the unrenormalized parameters of the avalanche picture \cite{thiery_microscopically_2017,de_roeck_stability_2017,luitz_how_2017,thiery_many-body_2018}.
        }
	\label{fig:nimofgETH}
\end{figure} 

\subsection{Re-entrance in interaction  strength}

\begin{figure} 
  \raggedleft
  \includegraphics[width=1.1\linewidth]{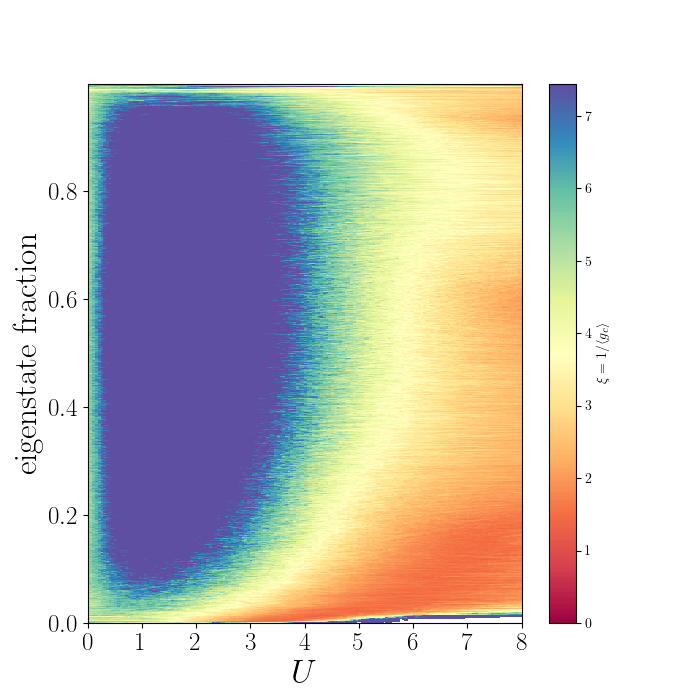}
  \caption{\textbf{Re-entrance of the localized phase as a function of interaction strength:}
    Localization length (extracted from critical tilt,
    $\xi = [\bar g_c]^{-1}$)
    as a function of eigenstate fraction and interaction strength $U$
    for $N = 12$-site disordered chains with width $W = 2$,
    at half-filling.
    See Fig.~\ref{fig:phdU6o12} for localization length as a function of energy.
    There is an intermediate delocalized regime between $U \approx 0.5-4$.
    Strong repulsion in the large $U$ limit freezes the system\cite{seetharam_absence_2018}; cf the work of Giudici et al. on lattice gauge theories, in which confinement plays a similar role \cite{giudici_breakdown_2020}.
    The low-energy delocalized regime for $U \gg 1$ may be related to the model's quantum phase transition at the isotropic point $U = 2$.
    See Fig.~\ref{fig:phdU6o12-e} for the same data plotted against energy.
  }
  \label{fig:phdU6o12} 
\end{figure}

Now fix the disorder width at $W = 2$ and vary the interaction strength $U$.
(We show the resulting localization lengths in Fig.~\ref{fig:phdU6o12}.)
At $U = 0$, the system is Anderson localized;
as $U$ increases we see the system delocalize (except near the band edges).
But for $U \gtrsim 3$ the system appears to localize once again.

This is a priori surprising.
In a na\"ive picture of Anderson localization with interaction added perturbatively,
we would expect the Anderson eigenstates to more strongly dephase and delocalize as we increase $U$;
in the more sophisticated avalanche picture, we would expect increasing interactions to increase the initial density of thermalized ``bare spots''.

But this reentrance is consistent with the work of Bera et al.\cite{bera_many-body_2015}
They study the random-field XXZ model, as we do;
they characterize MBL by probing the extent to which eigenstates of the many-body interacting Hamiltonian can be approximated by
Slater determinants of single-particle states.
They find (in their Fig.~1b) a reentrance in interaction very similar to ours.

\section{Conclusion}

In this manuscript we showed that an imaginary vector potential can provide direct access to localization lengths of noninteracting as well interacting localized systems. The crucial quantity is the ``critical tilt'', i.e., the vector potential at which an eigenenergy develops an imaginary part. We argue that the critical tilt measures the localization length of the underlying Hermitian system. We show this explicitly for the non-interacting limit with a lead site connecting the two ends of an open system. Importantly, we show that the localization length of an open disordered chain is given directly by the critical tilt (or critical imaginary flux per bond) of a ring made of the open chain plus a tunneling site.
We then argue that the connection remains for ordinary periodic boundary conditions, and that interactions cause a ``broadening'' in the appearance of imaginary eigenvalues. Finally, we use the correspondence to extract the localization length most relevant for transport properties in interacting, many-body localized, systems.

By using the critical tilt to measure localization length, we study the MBL transition.
We find re-entrance in the interaction strength $U$, which is a priori surprising
but consistent with prior work\cite{bera_many-body_2015},
and with the MBL transition found by Giudici et al. \cite{giudici_breakdown_2020} in $U(1)$ lattice gauge theories, where 1D Coulomb interactions cooperate with disorder to localize the system, rather than competing.
We also find a critical exponent $\nu\approx 1$, in agreement with other studies of the critical length exponent \cite{ }. 

Our work has finite overlap with the work by Hamazaki et al.\cite{hamazaki_non-hermitian_2019}, which studies directly the disorder-tuned PT breaking transition of a disordered system with a finite tilt. We emphasize the difference between our works: We are seeking to characterize the tilt-free MBL transition, whereas Ref. \onlinecite{hamazaki_non-hermitian_2019} studies the finite tilt transition and obtain a critical length exponent of $\nu=1/2$. Indeed, our results suggest that the two transitions---finite tilt and zero-tilt---are in different universality classes, as we confirm earlier observations on finite systems of $\nu=1$ at the transition.  This raises the possibility that the non-Hermitian system could provide differentiation between the different length scales explored, e.g., in Ref. \onlinecite{Vadim-scales}.

\begin{figure*}
  \begin{minipage}{0.45\textwidth}
    \includegraphics[width=\linewidth]{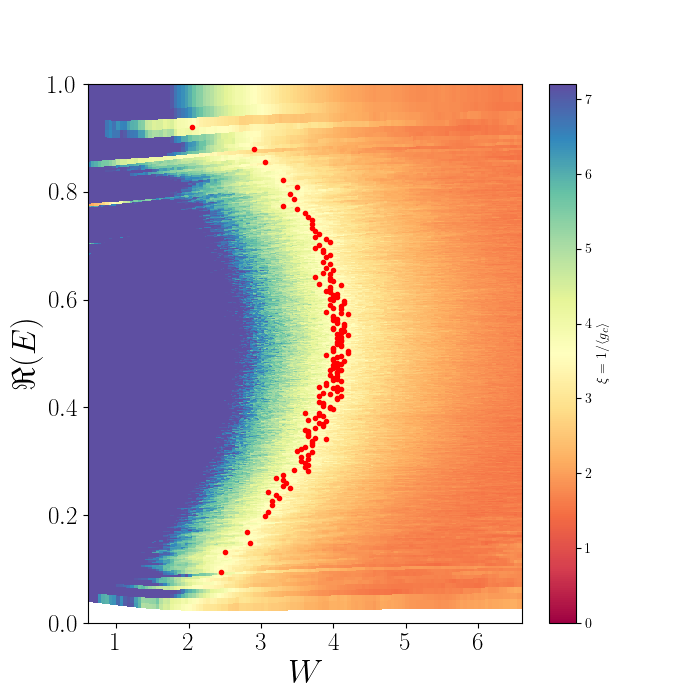}
  \end{minipage}
  \begin{minipage}{0.45\textwidth}
    \includegraphics[width=\linewidth]{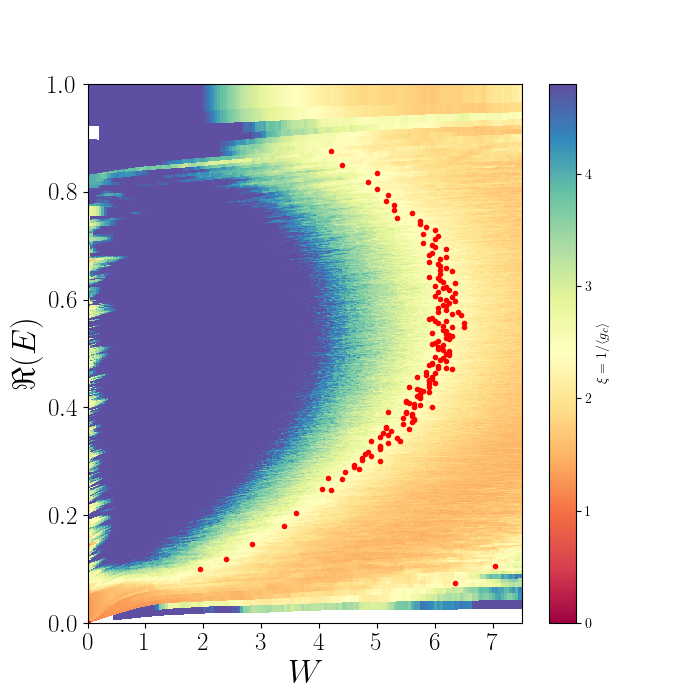}
  \end{minipage}
  
    \caption{
      \textbf{Phase diagram of the disordered, interacting Hamiltonian \eqref{eq:ham} }
      at
      system size $N = 12$,
      interaction strength $U = 1$ (\textbf{left}) and $U = 3$ (\textbf{right}),
      half filling,
      extracted from critical tilts.
      The color scale is $\xi \equiv [\bar g_c]^{-1}$;
      $x$ axis is still disorder width $W$;
      $y$ axis is now energy (rescaled by bandwidth);
      cf Fig.~\ref{fig:phd6o12}.
      Red dots indicate $\xi = 0.3N$ ($U = 1$, left) or $\xi = 0.2N$ ($U = 3$, right),
      which is consistent with the crossing in the scaling collapse of Fig. \ref{fig:scale}.
    }
  \label{fig:phd6o12-e}
\end{figure*}
\begin{figure}
  \includegraphics[width=\linewidth]{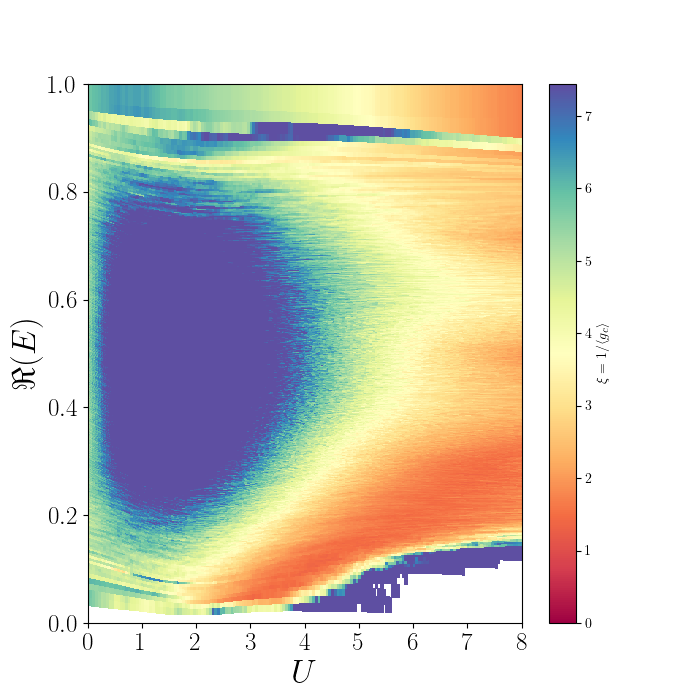}
  \caption{\textbf{Re-entrance of the localized phase as a function of interaction strength and energy:}
    Localization length (extracted from critical tilt,
    $\xi = [\bar g_c]^{-1}$)
    as a function of eigenstate energy (rescaled by bandwidth) and interaction strength $U$
    for $N = 12$-site disordered chains with width $W = 2$,
    at half-filling
    (cf Fig.~\ref{fig:phdU6o12}).
    There is an intermediate delocalized regime between $U \approx 0.5-4$.
    Strong repulsion in the large $U$ limit freezes the system\cite{seetharam_absence_2018}; cf the work of Giudici et al. on lattice gauge theories, in which confinement plays a similar role \cite{giudici_breakdown_2020}.
    The low-energy delocalized regime for $U \gg 1$ may be related to the model's quantum phase transition at the isotropic point $U = 2$.
  }
  \label{fig:phdU6o12-e} 
\end{figure}

Our per-eigenstate critical tilt measures a localization length of each eigenstate.
It could be recast as the critical tilt in each energy window, $[E, E + \delta E)$, as is done in Figs. \ref{fig:phd6o12-e} and \ref{fig:phdU6o12-e} in App. \ref{app-e}. This is in some sense the MBL-side mirror image of the slow thermalization rates measured by Pancotti et al.\cite{pancotti_almost-conserved_2018} on the ETH side of the transition.
Pancotti et al. characterize the distribution of operator decay rates of the most nearly conserved local operators in terms of extreme value statistics;
these anomalously slow decay rates probe the \emph{least thermal} states on the ETH side of the MBL transition---those states that take the longest to decay to equilibrium.
As disorder increases they find a crossover from tight Gumbel statistics to heavy-tailed Fr\'echet statistics.
Measuring critical tilt in an energy window, by contrast, would measure the localization lengths of the \emph{least localized} states on the MBL side of the transition.
It would be interesting to characterize the distribution of $g_c$ across disorder realizations in terms of extreme value statistics. This would be the subject of future work. 

It is also interesting to consider the critical tilt in light of the avalanche picture of De Roeck et al.
\cite{thiery_microscopically_2017,de_roeck_stability_2017,luitz_how_2017,thiery_many-body_2018}.
In the avalanche narrative, one adds interactions to an Anderson insulator via a quasi-perturbative RG scheme;
regions where interactions cannot be treated perturbatively are treated as thermal inclusions.
They take the microscopic system to be parametrized by two parameters,
an Anderson localization length and a density of these initial thermal inclusions. This is the basis for the RG picture in Ref. \onlinecite{morningstar_renormalization-group_2019}.
In this picture it is not enough to consider the critical $g_c$ in some energy window:
this corresponds (we expect) to the localization length of the \emph{least localized} eigenstate.
But a single delocalized eigenstate should not be enough to destabilize a surrounding localized region.
Rather, one needs a finite fraction of eigenstates to be delocalized.
In this picture $g_{\ff}$ (for systems of some fixed size $N$) corresponds to the localization length that is the key variable in the avalanche picture RG flow;
in principle, computing $g_{\ff}$ as a function of system size will allow one to probe the flow of that variable,
providing a sensitive test of the avalanche picture.
Because---for open boundary conditions---eigenstates of the tilted system are gauge-equivalent to eigenstates of the underlying Hermitian system, 
tensor network techniques\cite{pollmann_efficient_2016,khemani_obtaining_2016,wahl_efficient_2017,yu_finding_2017,devakul_obtaining_2017,chandran_spectral_2015} may give access to these quantities for large systems.

The finite-fraction tilt $g_{\ff}$ will also probe the unrenormalized ``bare spot'' probability:
that is, the probability that a subsystem will be thermal.
Recall (Fig.~\ref{fig:nimofgETH}) that the distribution of $g_c$ extends all the way to zero,
even for large disorder width.
This is because some (anomalous) disorder realizations have eigenstates stretching across the system.
If a particular disorder width has small $g_\ff$, it is effectively thermal---that is, it is a ``bare spot'', in the language of the avalanche picture.
More careful measurements of the distribution of $g_c$, and the analogous distribution for $g_{\ff}$, at small system size will therefore also characterize the unrenormalized,
microscopic inputs into the avalanche picture.

\begin{acknowledgements}
  We thank Bernd Rosenow, Sarang Gopalakrishnan, and Vadim Oganesyan for many helpful conversations; we also thank Naomichi Hatano and an anonymous reviewer for commentary that prompted us to sharpen our understanding and arguments.
  
GR is grateful for funding from NSF grant 1839271 as well as to the Simons Foundation, the Packard Foundation, and the IQIM, an NSF frontier center partially funded by the Gordon and Betty Moore Foundation. The authors thank FAU Erlangen-N\"urnberg's Prof. Dr. Kai P. Schmidt for setting up and accompanying the team of researchers involved in this work. We gratefully acknowledge funding received by the German Academic Exchange Service.
This work is partially supported by the U.S. Department of Energy (DOE), Office of Science, Office of Advanced Scientific Computing Research (ASCR) Quantum Computing Application Teams program, under fieldwork proposal number ERKJ347.
\end{acknowledgements}

\bibliography{references.bib}

\appendix


\begin{figure}
  \begin{minipage}{0.39\textwidth}
    \includegraphics[width=\textwidth]{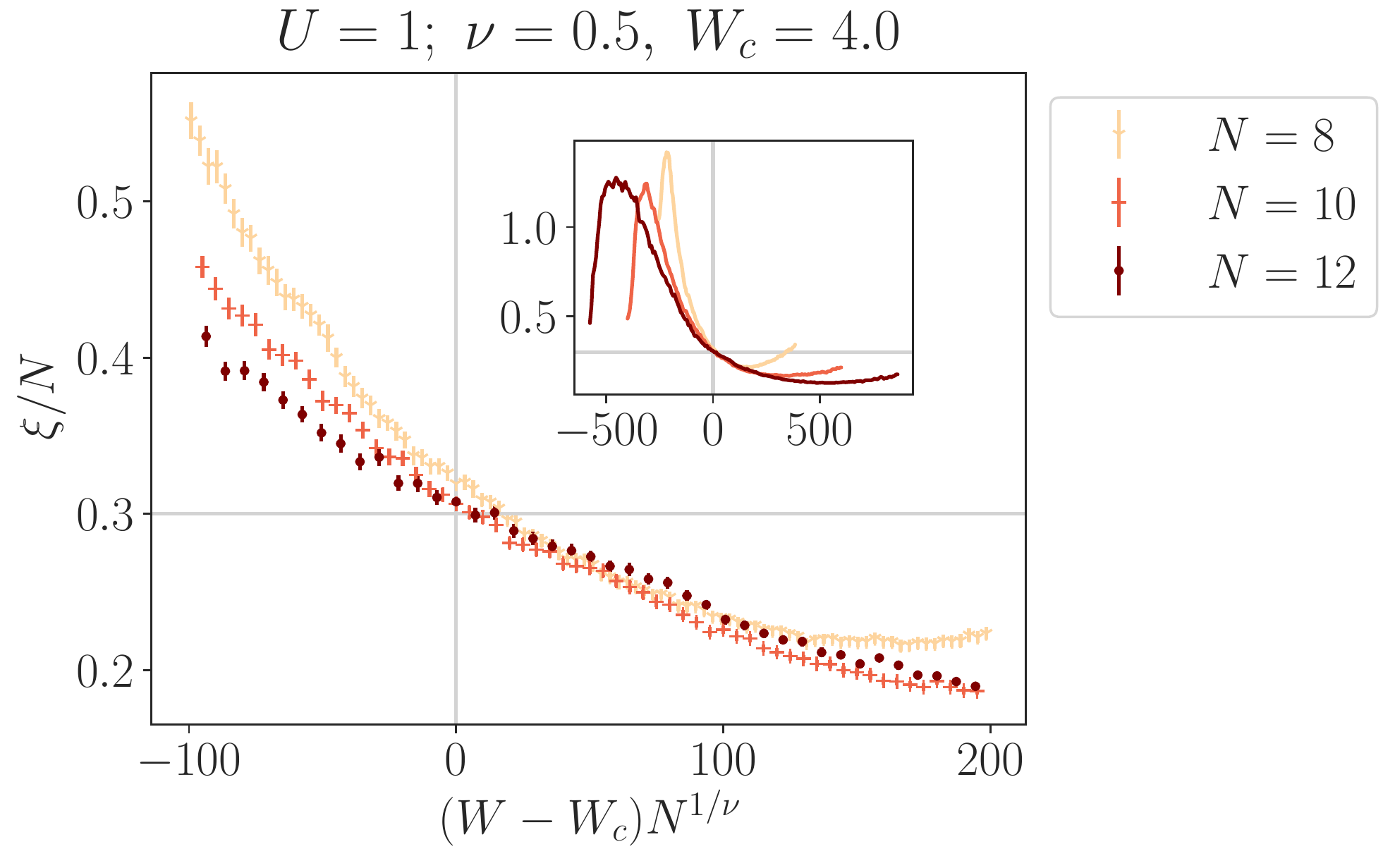}
  \end{minipage}
  
  \begin{minipage}{0.39\textwidth}
    \includegraphics[width=\textwidth]{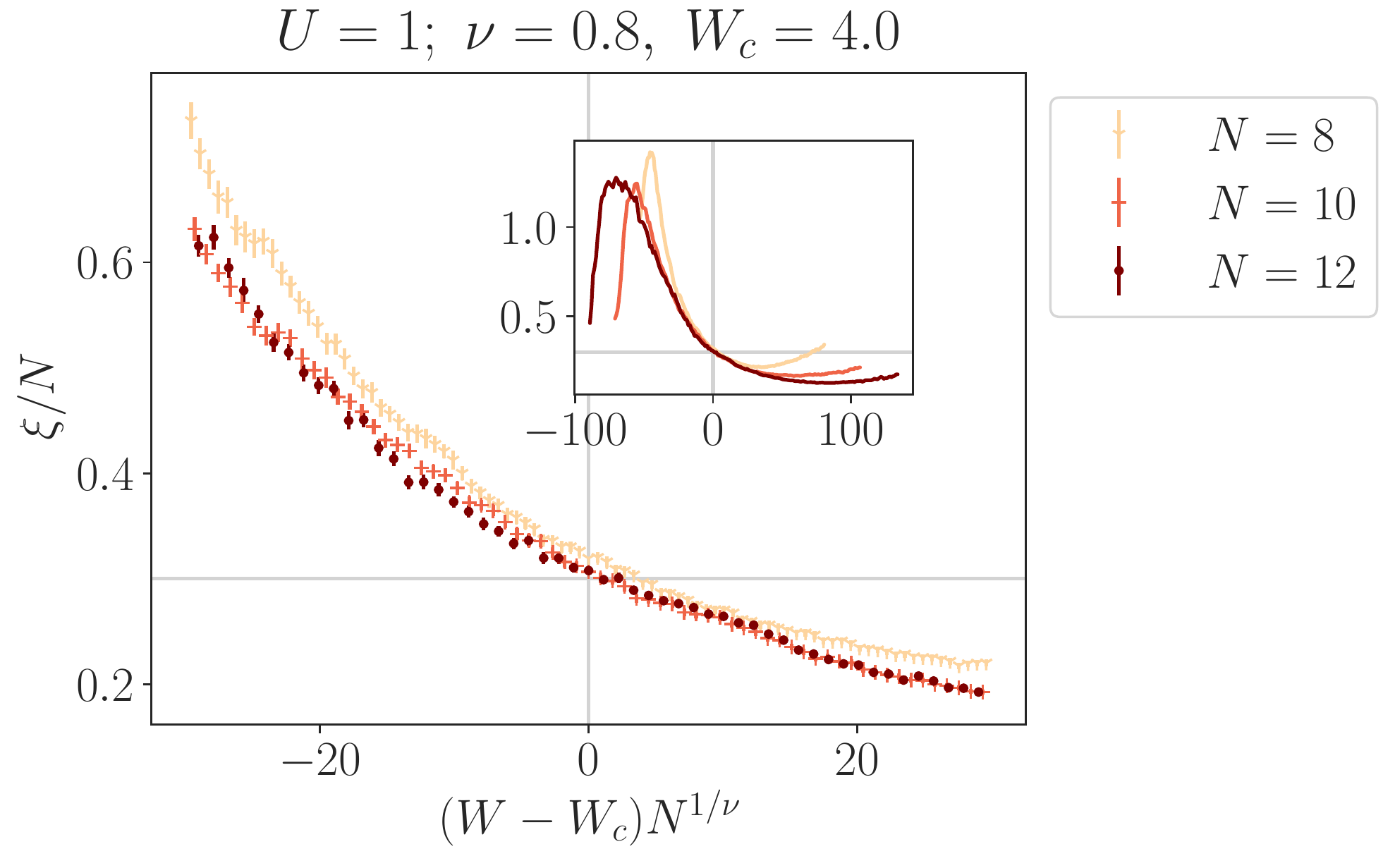}
  \end{minipage}
  
  \begin{minipage}{0.39\textwidth}
    \includegraphics[width=\textwidth]{NHMBL23_scaling-inset-U1-nu1-0-Wc4-0.pdf}
  \end{minipage}
  
  \begin{minipage}{0.39\textwidth}
    \includegraphics[width=\textwidth]{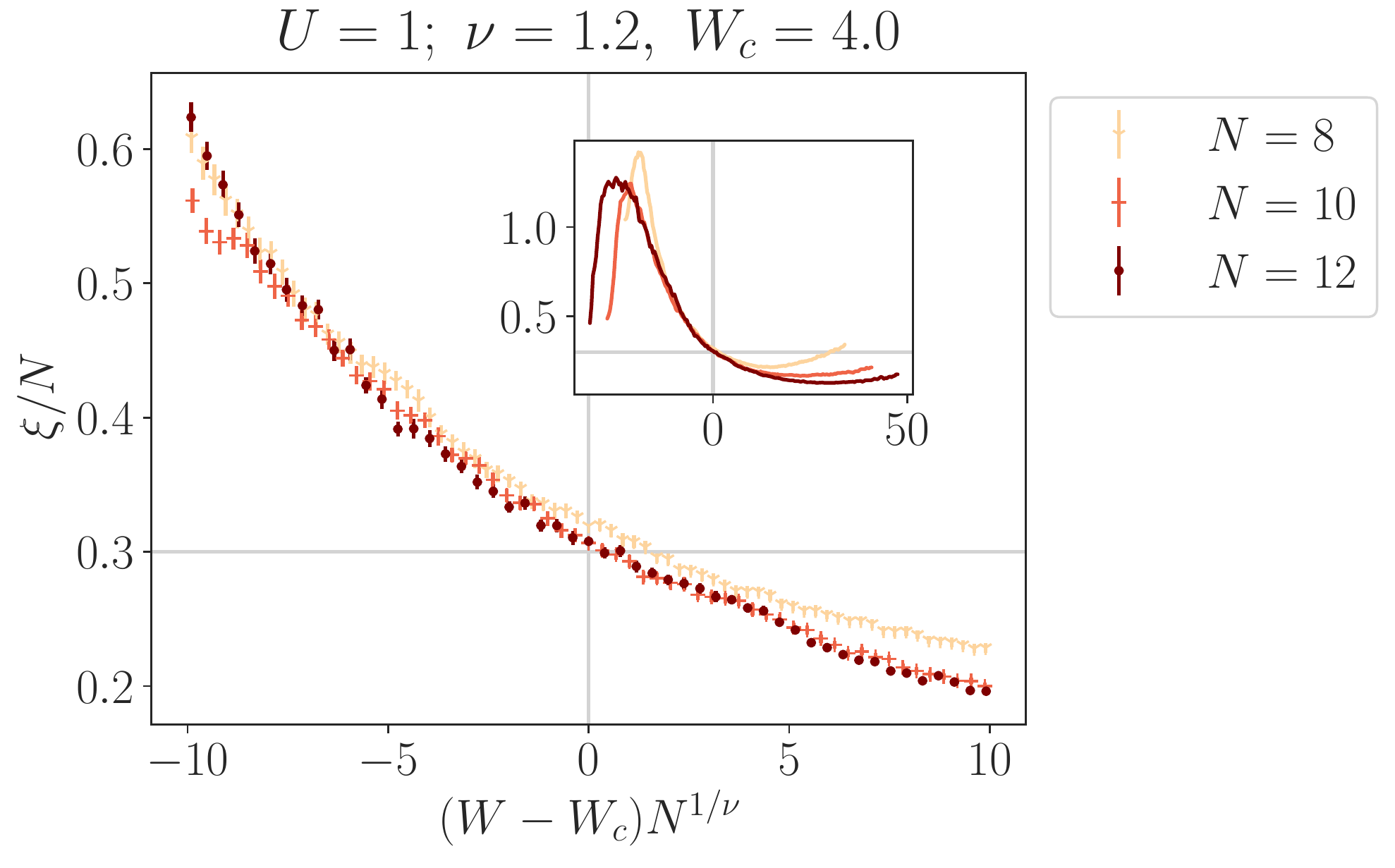}
  \end{minipage}
  
  \begin{minipage}{0.39\textwidth}
    \includegraphics[width=\textwidth]{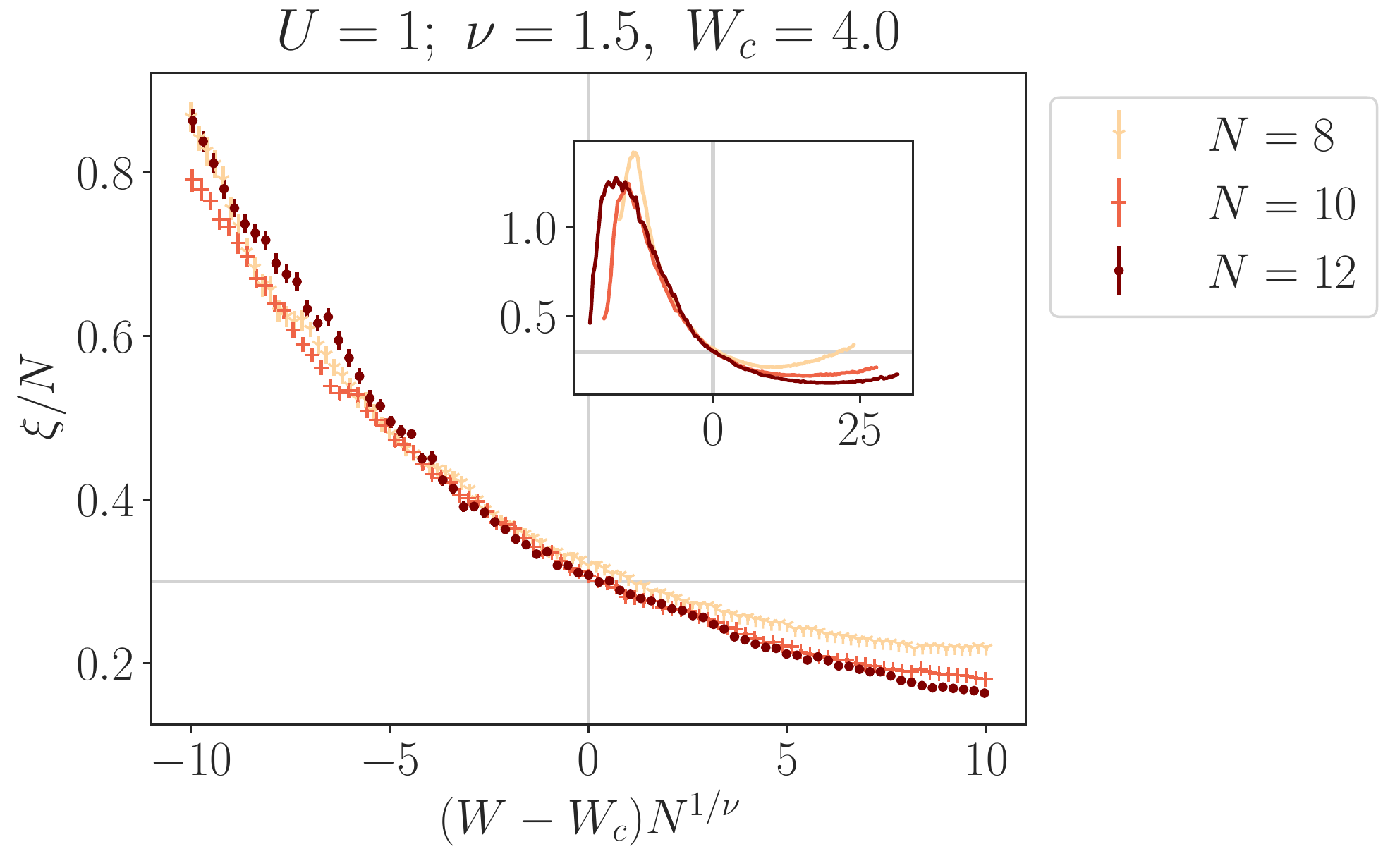}
  \end{minipage}
  
 \caption{Putative finite-size scaling for $U = 1, W_c = 4$, and a variety of $\nu$.}
  \label{fig:U1-nu}
\end{figure}


\begin{figure}[t!]
  \begin{minipage}{0.40\textwidth}
    \includegraphics[width=\textwidth]{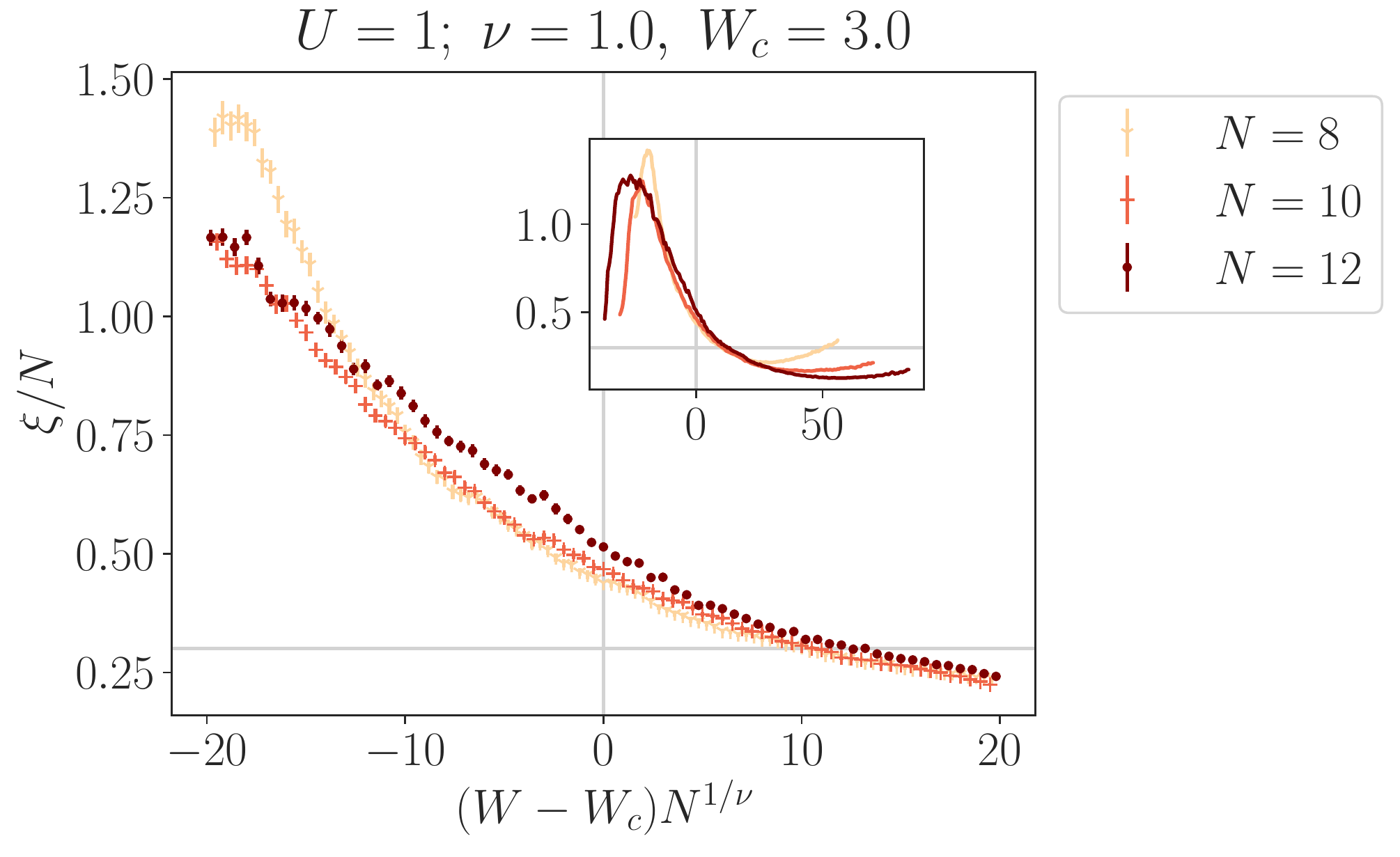}
  \end{minipage}
  
  \begin{minipage}{0.40\textwidth}
    \includegraphics[width=\textwidth]{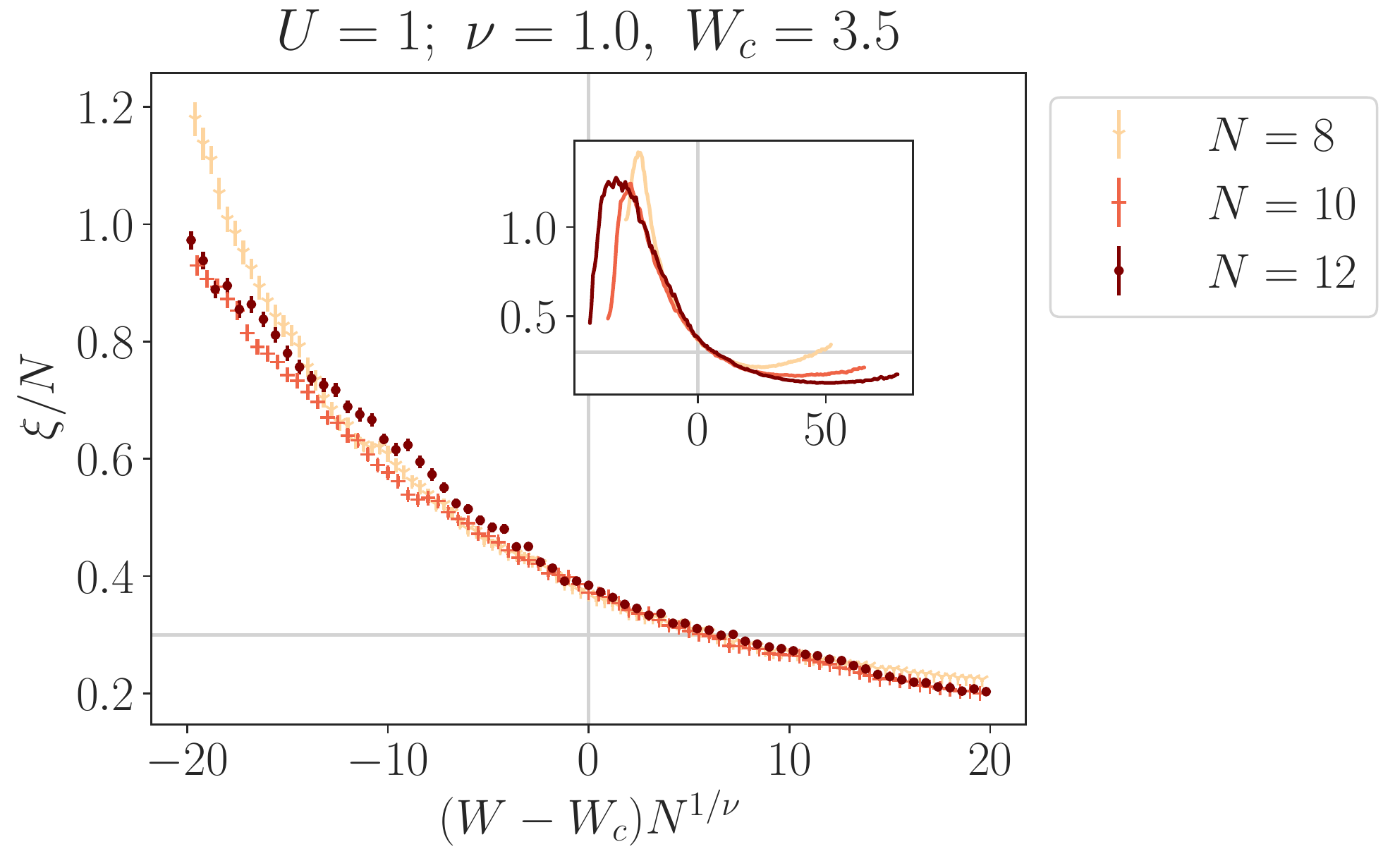}
  \end{minipage}

  \begin{minipage}{0.40\textwidth}
    \includegraphics[width=\textwidth]{NHMBL23_scaling-inset-U1-nu1-0-Wc4-0.pdf}
  \end{minipage}
  
  \begin{minipage}{0.40\textwidth}
    \includegraphics[width=\textwidth]{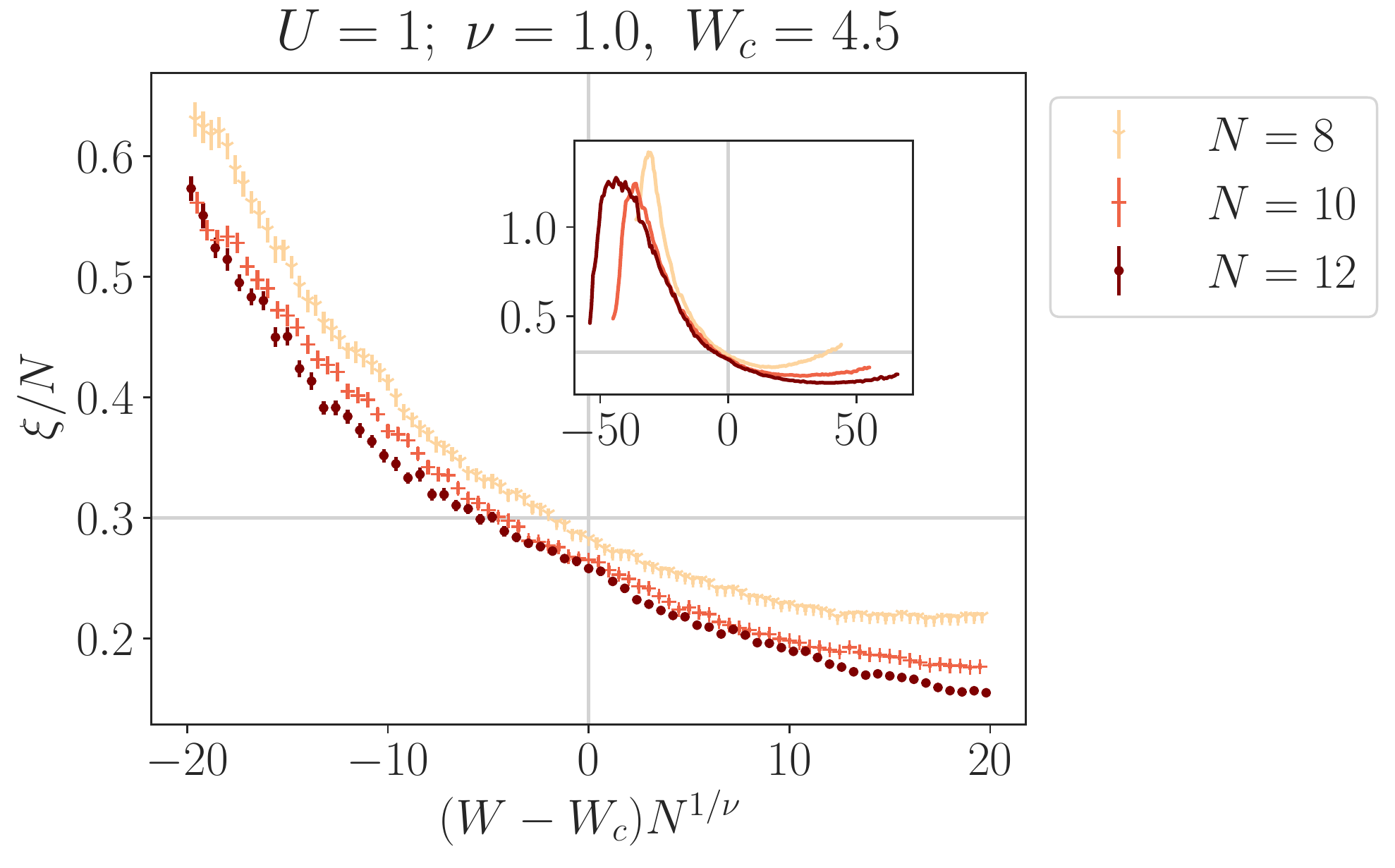}
  \end{minipage}
  
  \caption{Putative finite-size scaling for $U = 1, \nu = 1.0$, and a variety of $W_c$.}
\end{figure}

\section{Jordan-Wigner tranform}
For convenience we note that the Hamiltonian \eqref{eq:ham} has Jordan-Wigner transform
\begin{align}
  \begin{split}
    H &= t \sum_j [e^gc^\dagger_j c_{j + 1} + e^{-g}c^\dagger_{j+1}c_j] \\
    &\qquad + U \sum_j n_j n_{j+1} + \sum_j h_j n_j 
  \end{split}\\
  \begin{split}
      &= t \sum_j [e^g S^+_j S^-_{j+1} + e^{-g} S^+_{j+1} S^-_j] \\
      &\qquad + U \sum_j S^z_j S^z_{j+1} + \sum_j h_j n_j 
  \end{split}\\
  \begin{split}
      & \propto t \sum_j \frac 1 2 [e^g S^+_j S^-_{j+1} + e^{-g} S^+_{j+1} S^-_j] \\
      &\qquad + \frac 1 2 U \sum_j S^z_j S^z_{j+1} + \sum_j \frac 1 2 h_j S^z_j
  \end{split}
\end{align}
which in the $g = 0$ case reduces to
\begin{align}
  \begin{split}
  H &\propto \sum_j \left[t (S^x_j S^x_{j+1} + S^y_{j} S^y_{j+1}) + \frac U 2 S^z_j S^z_j\right] \\
  &\qquad + \sum_j \frac {h_j} 2 S^z_j
  \end{split}
\end{align}
(hence the choice of factors of 2).

\section{Phase diagrams as a function of energy \label{app-e}}

In Fig. \ref{fig:phd6o12} we plotted the localization length $\xi$ (extracted from the tilt) as a function of disorder width and the ``eigenstate fraction''---where in the sorted list of eigenstates a particular eigenstate falls.
In Figs \ref{fig:phd6o12-e}, \ref{fig:phdU6o12-e} we plot the localization length $\xi$ as a function of energy (normalized by the bandwidth of each disorder realization).
To be more precise, we
\begin{enumerate}
\item average the energies for eigenstate $\alpha \in 1 \dots {N \choose N/2}$ (at fixed disorder width $W$ and interaction strength), and
\item average the critical tilt $g_c$ for eigenstate $\alpha$ and extract the localization length.
\end{enumerate}
We plot the localization length (averaged in this sense) against the disorder width $W$ or interaction strength $U$ and the energy (averaged in this sense).

This changes the shape of the phase diagram, because the density of states is heuristically
\begin{equation}
  \rho(E) \propto \frac 1 {\sigma\sqrt{2\pi N} } e^{-E^2/2\sigma^2\sqrt{N} },\quad \sigma \sim \sqrt{N(t^2 + W_c^2)}
\end{equation}
(before bandwidth normalization).

The rescaling highlights certain ``glitches'' (e.g. in Fig.~\ref{fig:phd6o12-e} near $\Re(E) \approx 0.9 \text{[bandwidth]}$).
These also appear in Fig \ref{fig:phd6o12} but they are almost imperceptible because they only span one or two states.

We suspect that the glitches result from the presence or absence of resonances in the particular disorder realizations we use.
At very low or very high energy density, eigenenergies are widely spaced,
so interactions are less likely to link subsystems and cause them to dephase each other.
This is why the system is (at finite size) more localized near the edge of the spectrum.
But if---due to the vagaries of a particular disorder realization---two subsystems have eigenenergies near the edge of the spectrum that line up, or disagree more than usual, they will be anomalously delocalized or localized.

These edge-of-spectrum effects are, strictly speaking, outside the scope of this work: they are likely the result of infinite-randomness ground state physics, rather than many-body localization properly understood.
(One can already glimpse a similarity to Dasgupta-Ma\cite{PhysRevB.22.1305} real-space renormalization group arguments in our explanation above.)

\section{Finite-size scaling}
\label{a:scale}

\begin{figure}[p]
  \begin{minipage}{0.40\textwidth}
    \includegraphics[width=\textwidth]{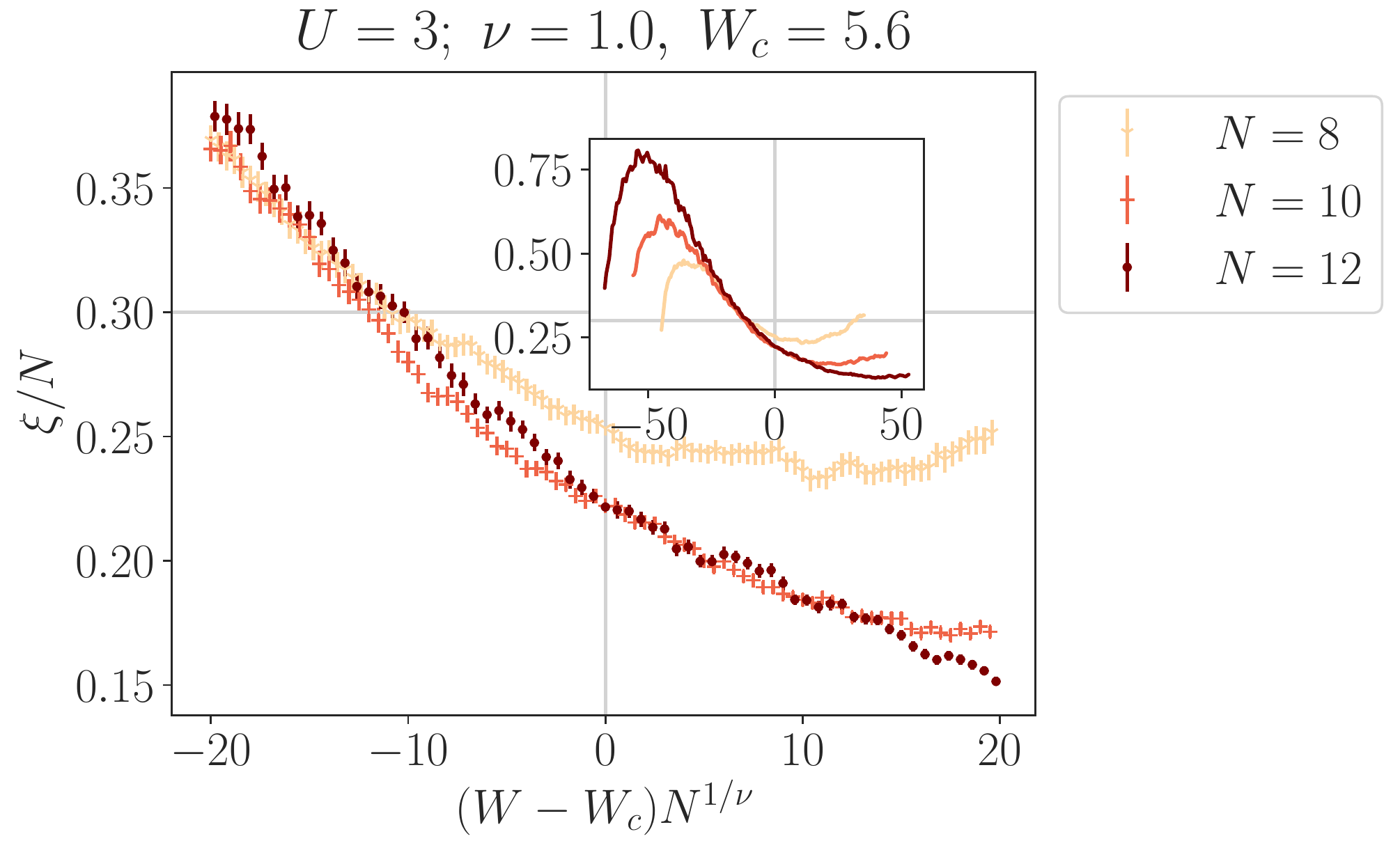}
  \end{minipage}
  
  \begin{minipage}{0.40\textwidth}
    \includegraphics[width=\textwidth]{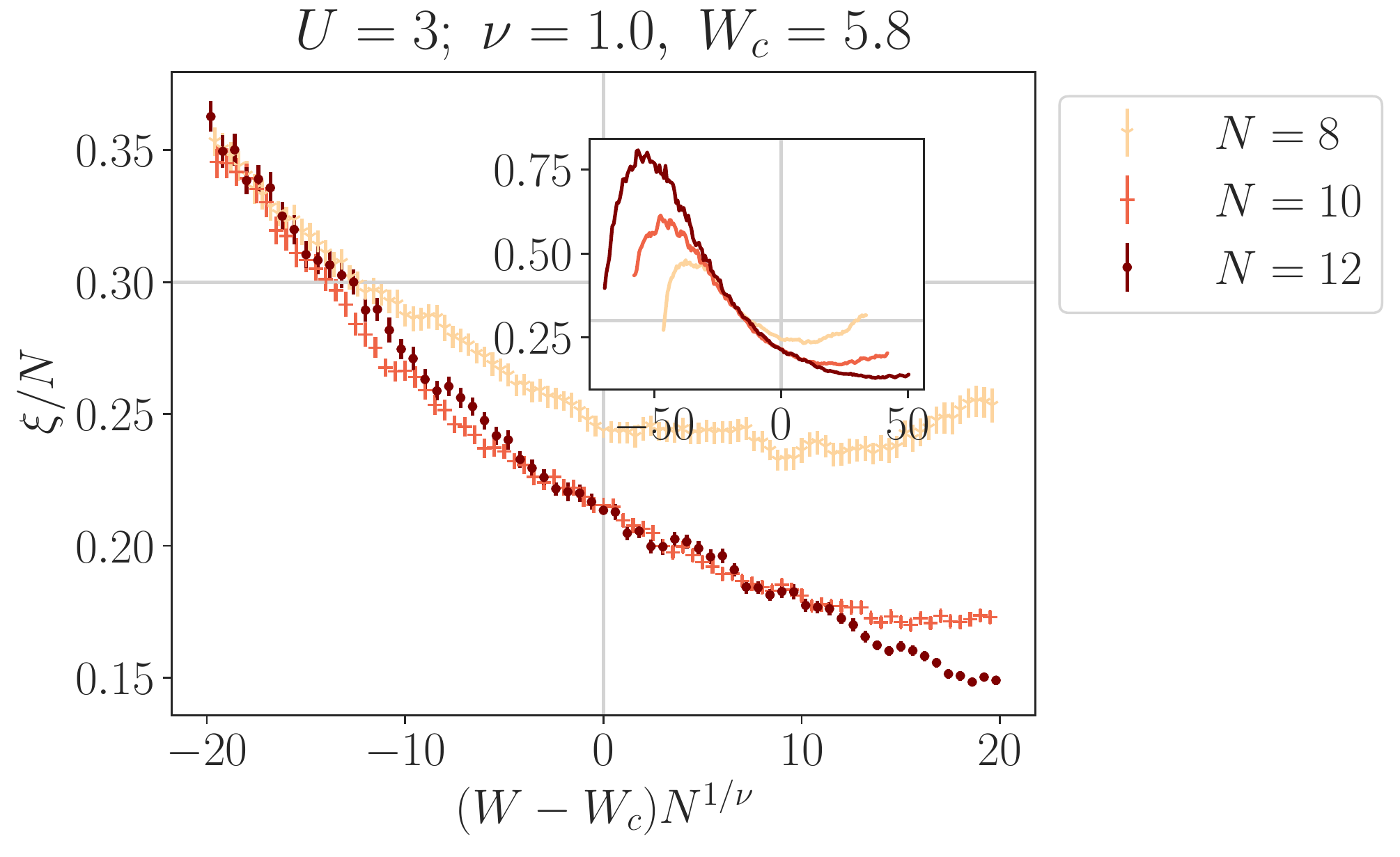}
  \end{minipage}

  \begin{minipage}{0.40\textwidth}
    \includegraphics[width=\textwidth]{NHMBL23_scaling-inset-U3-nu1-0-Wc6-0.pdf}
  \end{minipage}
  
  \begin{minipage}{0.40\textwidth}
    \includegraphics[width=\textwidth]{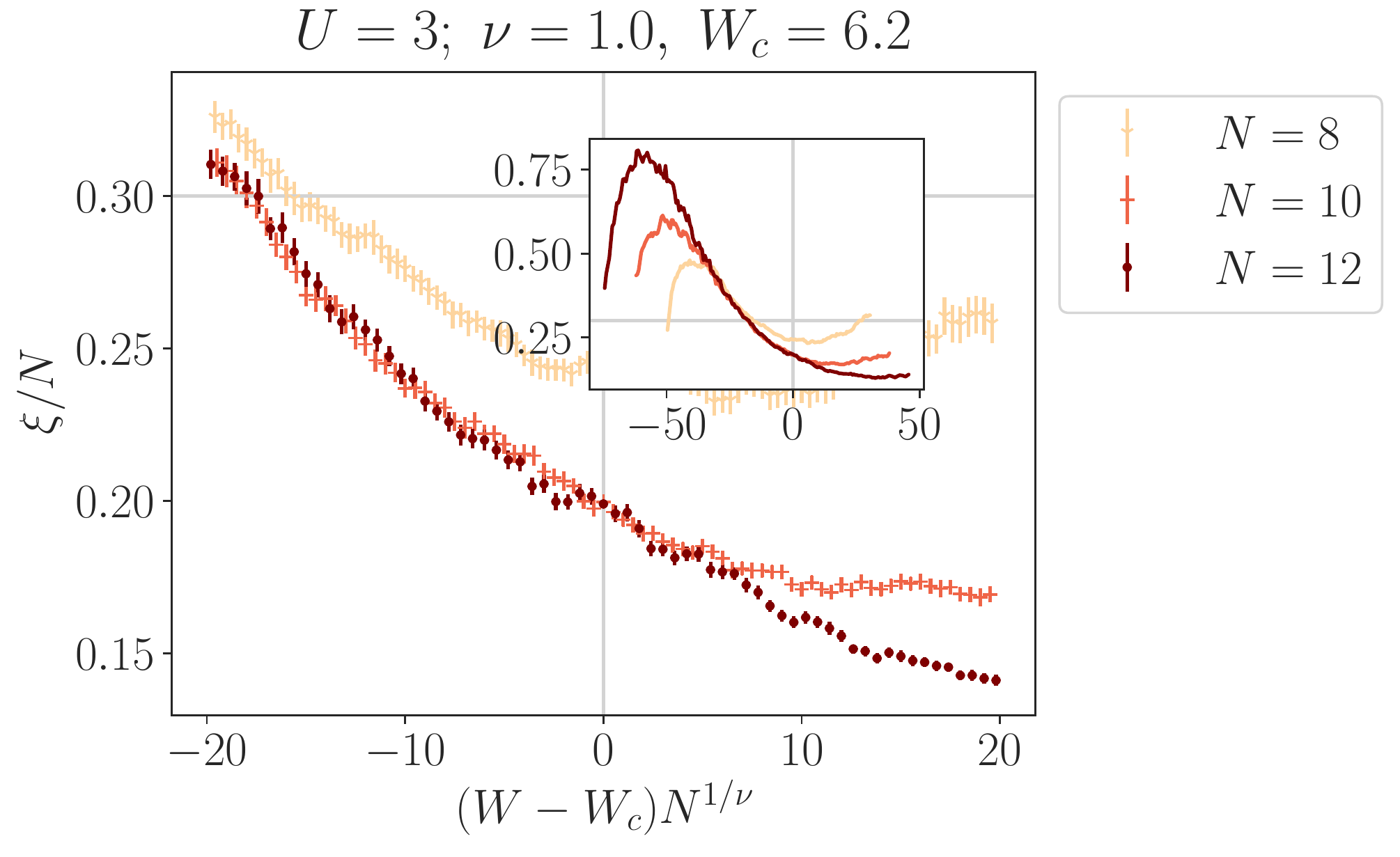}
  \end{minipage}
  
  \begin{minipage}{0.40\textwidth}
    \includegraphics[width=\textwidth]{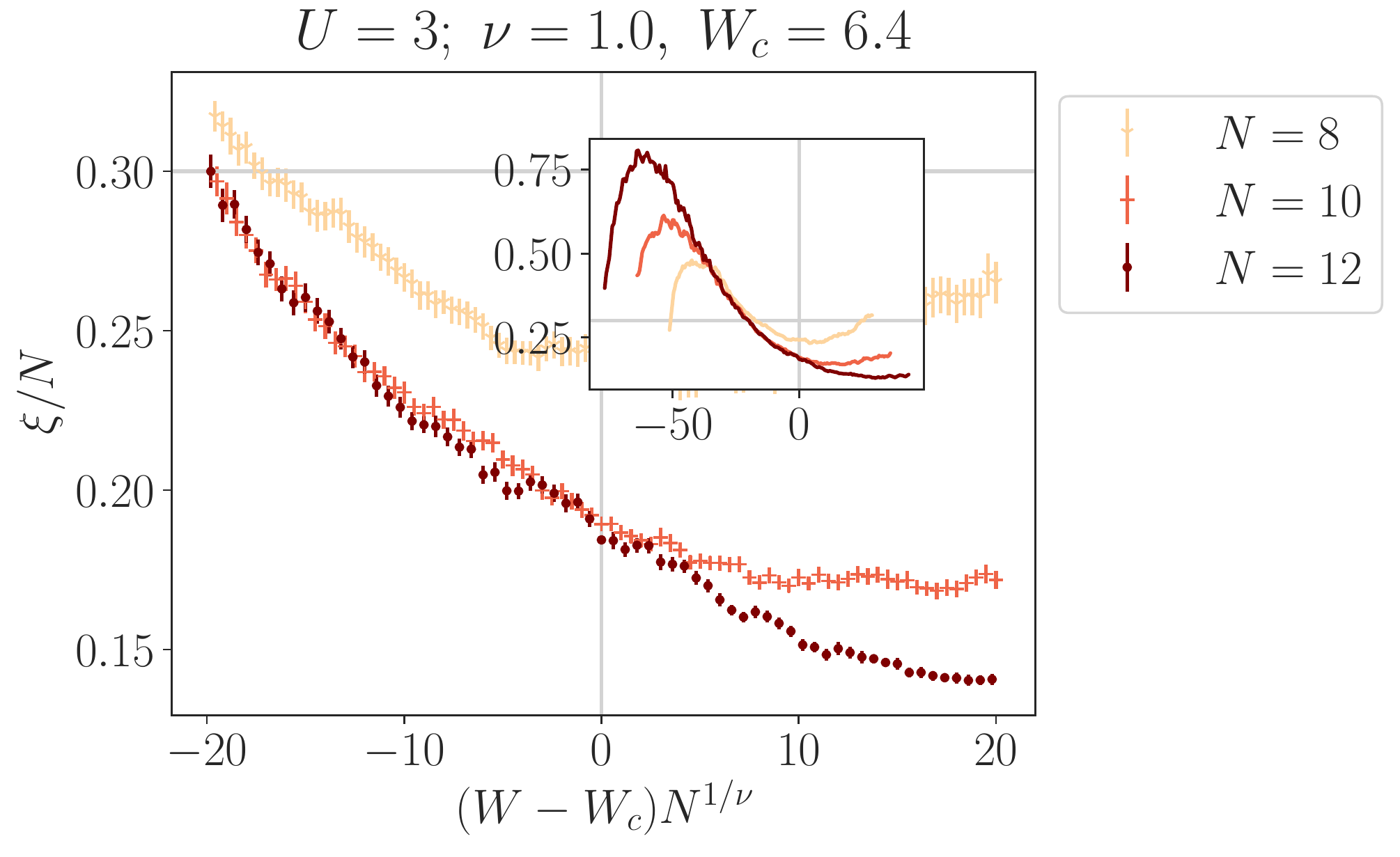}
  \end{minipage}
  
  \caption{Putative finite-size scaling for $U = 3, \nu = 1.0$, and a variety of $W_c$}
\end{figure}

\begin{figure}[p]
  \begin{minipage}{0.40\textwidth}
    \includegraphics[width=\textwidth]{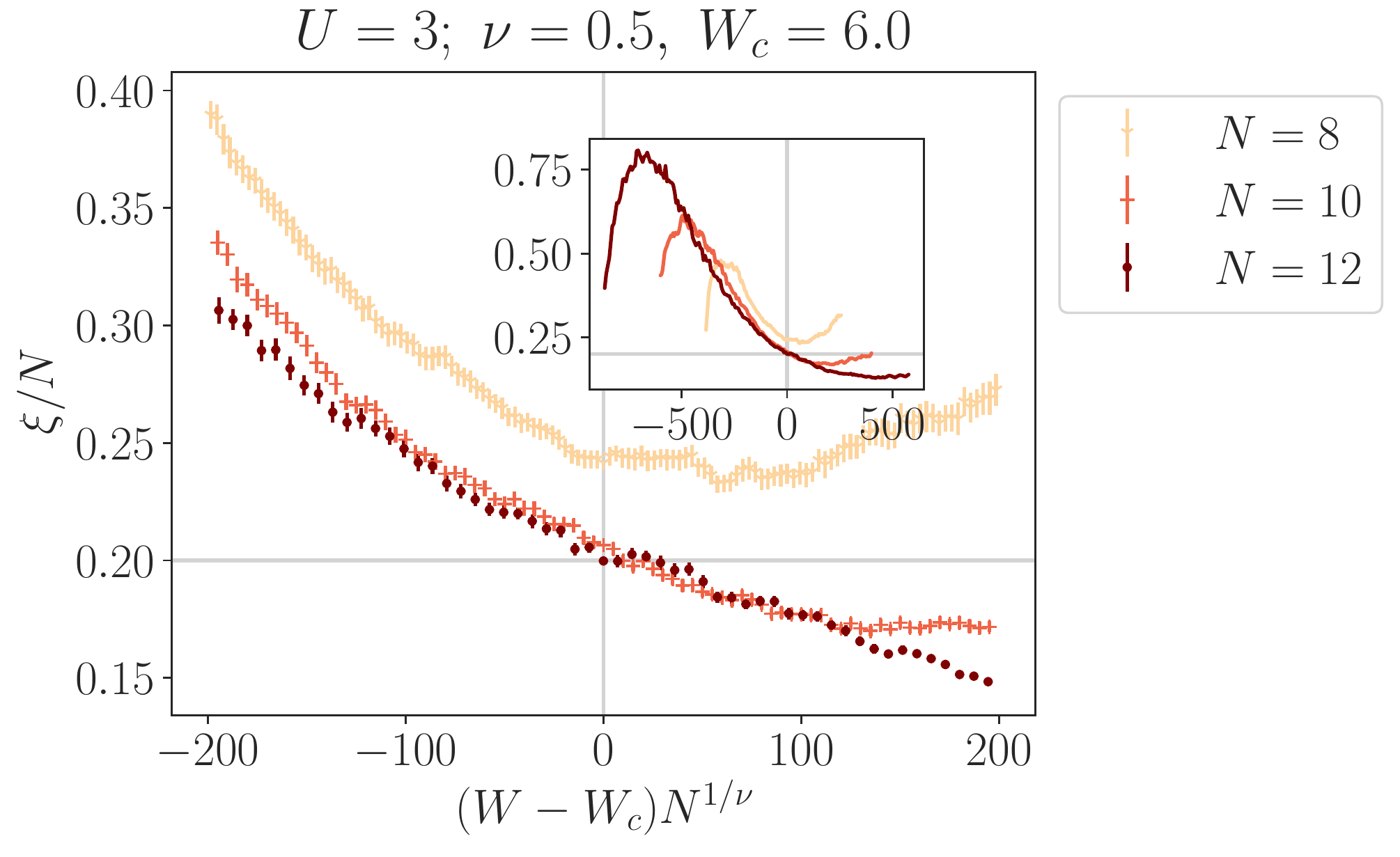}
  \end{minipage}
  
  \begin{minipage}{0.40\textwidth}
    \includegraphics[width=\textwidth]{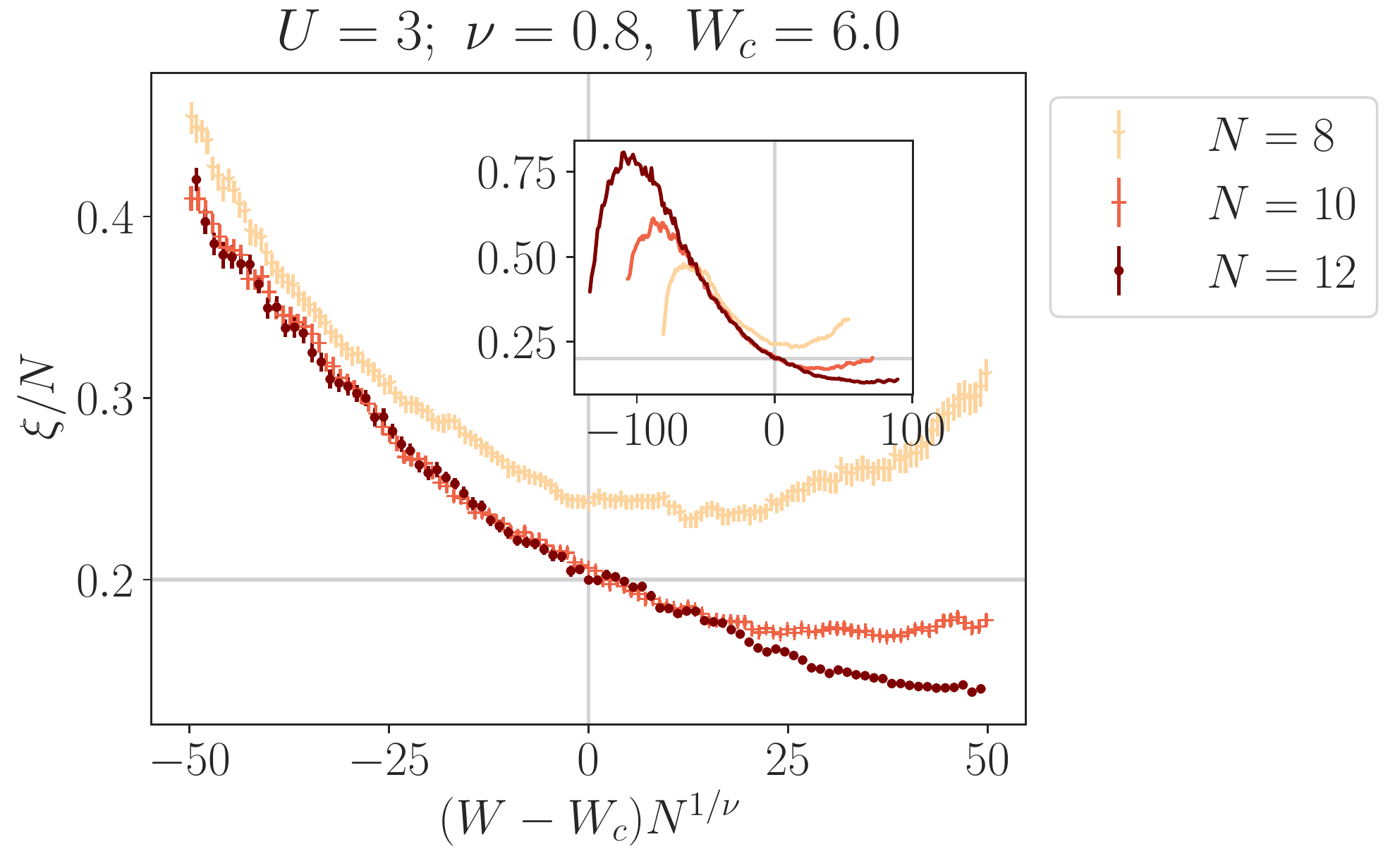}
  \end{minipage}
  
  \begin{minipage}{0.40\textwidth}
    \includegraphics[width=\textwidth]{NHMBL23_scaling-inset-U3-nu1-0-Wc6-0.pdf}
  \end{minipage}
  
  \begin{minipage}{0.40\textwidth}
    \includegraphics[width=\textwidth]{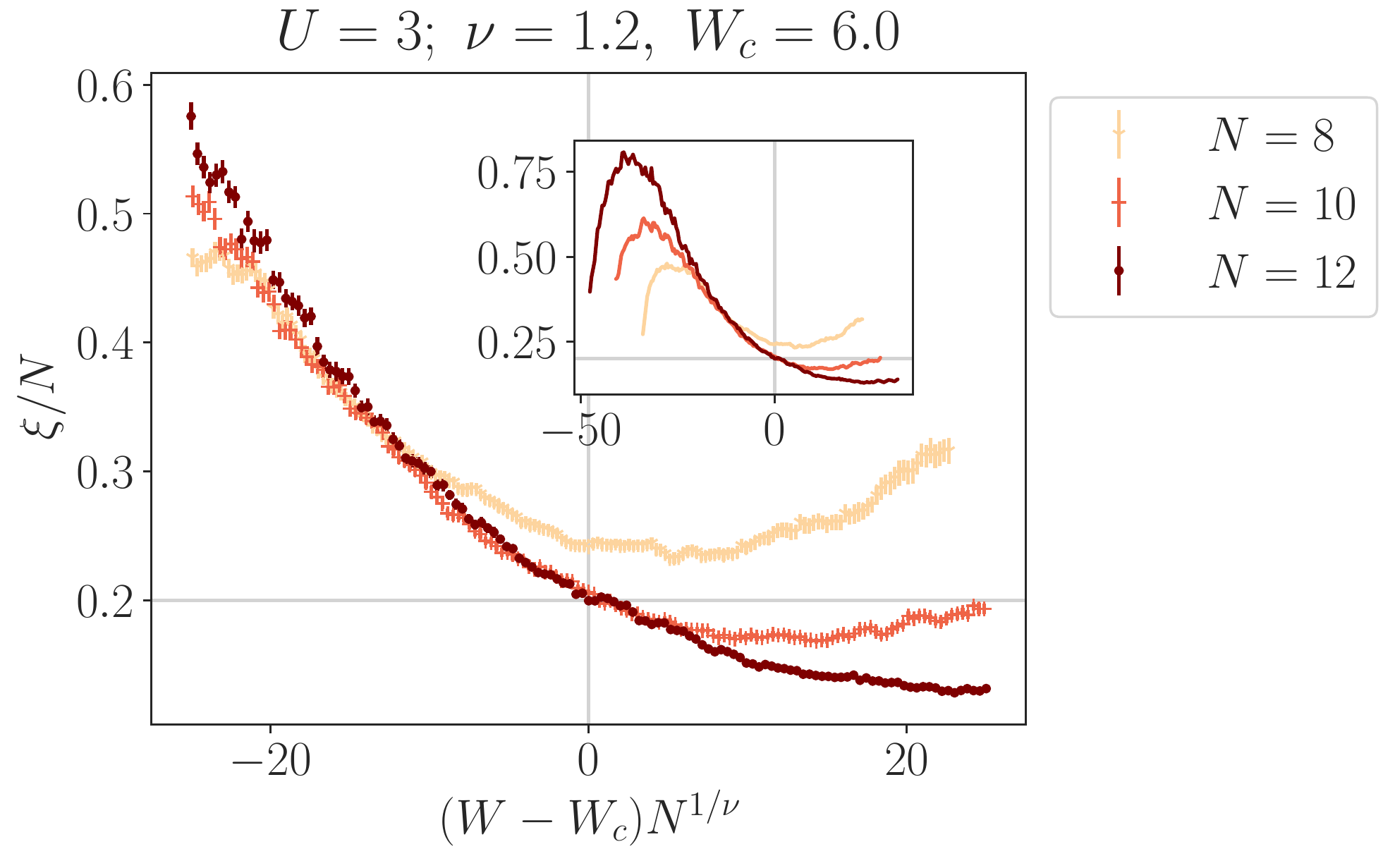}
  \end{minipage}
  
  \begin{minipage}{0.40\textwidth}
    \includegraphics[width=\textwidth]{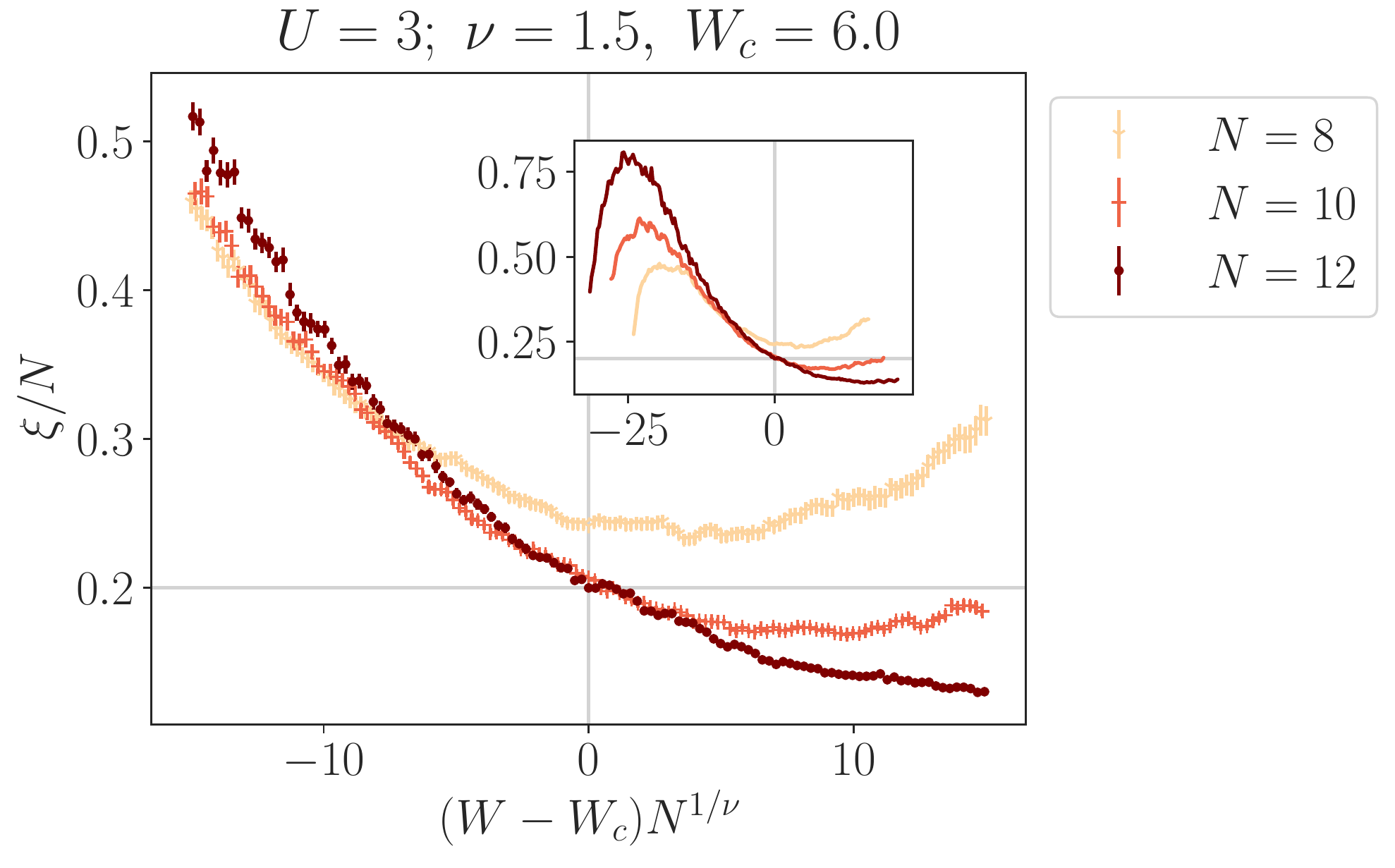}
  \end{minipage}
  
  \caption{Putative finite-size scaling for $U = 3, W_c = 6$, and a variety of $\nu$}
  \label{fig:U3-Wc}
\end{figure}

In Sec.~\ref{s:phase:iso} we claimed that our finite-size scalings gave $W_c = 4, 6$ for $U = 1,3$ respectively,
and $\nu = 1$ for both interaction strengths.
In Figs~\ref{fig:U1-nu}-\ref{fig:U3-Wc} we show (putative) finite-size scalings for a variety of $W_c, \nu$, so the reader can judge the accuracy and precision of our claims.


\end{document}